\documentclass[11pt,a4paper]{article}
\usepackage{jheppub}

\usepackage{amsmath,amssymb,amsfonts}
\usepackage{graphicx}
\usepackage{bbm}
\usepackage[dvipsnames]{xcolor}
\usepackage[normalem]{ulem}

\usepackage[nameinlink]{cleveref}
\usepackage{acronym}

\usepackage{feynmp}
\definecolor{tealblue}{rgb}{0.21, 0.56, 0.63}
\hypersetup{colorlinks=true,allcolors = tealblue,linktocpage=true}

\renewcommand{\figurename}{{\textbf{Figure}}}
\renewcommand{\thefigure}{{\textbf{\arabic{figure}}}}
\renewcommand{\tablename}{{\textbf{Table}}}
\renewcommand{\thetable}{{\textbf{\arabic{table}}}}

\newacro{UV}[UV]{ultraviolet}
\newacro{IR}[IR]{infrared}
\newacro{QFT}[QFT]{quantum field theory}
\acrodefplural{QFT}{quantum field theories}
\newacro{EFT}[EFT]{effective field theory}
\acrodefplural{EFT}{effective field theories}
\newacro{FRG}[FRG]{Functional Renormalisation Group}
\newacro{RG}[RG]{renormalisation group}

\newcommand{\eg}{e.g.}
\newcommand{\ie}{i.e.}

\newcommand{\regulatoroperator}{\ensuremath{\mathfrak R}}
\newcommand{\mans}{\ensuremath{s}}
\newcommand{\mant}{\ensuremath{t}}
\newcommand{\manu}{\ensuremath{u}}

\allowdisplaybreaks

\title{Asymptotically (un)safe scattering amplitudes from scratch: a deep dive into the IR jungle}
\author{Benjamin Knorr\,\href{https://orcid.org/0000-0001-6700-6501}{\protect \includegraphics[scale=.07]{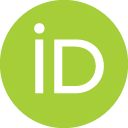}}\,}

\affiliation{Institut f\"ur Theoretische Physik, Universit\"at Heidelberg, Philosophenweg 12, 69120 Heidelberg, Germany}

\emailAdd{knorr@thphys.uni-heidelberg.de}

\abstract{We compute leading order quantum gravity contributions to a simple scalar scattering amplitude in Asymptotic Safety. Our model admits an analytic treatment so that several subtleties can be analysed. We find that (i) the existence of an asymptotically safe renormalisation group fixed point alone does not imply the boundedness of scattering amplitudes, (ii) gravitational logarithms can dominate the infrared regime of massless theories, (iii) a derivative expansion of the effective action fails quantitatively to predict the correct Wilson coefficients in massless theories, and (iv) standard renormalisation group improvement techniques fail qualitatively to describe the momentum dependence of correlation functions. Only momentum-dependent computations can resolve these issues. For theories that include massive fields, the derivative expansion can work effectively in most cases, but it can still fail for classically marginal couplings, and purely gravitational couplings. We also speculate about an effective realisation of the no-global-symmetries conjecture in Asymptotic Safety.}

\begin{document}
\maketitle

\section{Introduction}\label{sec:intro}

Deriving concrete predictions from quantum gravity is challenging~\cite{Buoninfante:2024yth}, much more so if the predictions ought to be analytical. In this paper, we provide such an example in the context of Asymptotic Safety. In this scenario, we assume that we can describe quantum gravity as a \ac{QFT} even at Planckian scales. Specifically, we will discuss leading quantum-gravity effects in the simplest-possible matter theory, that is a scalar field theory. More concretely, we will mostly consider the two-to-two scattering of two shift-symmetric scalar fields. This reduces the complexity since no potential is allowed, but we will also discuss the effect of mass terms in a second step. Moreover, we will assume that we can probe the theory in a weak-gravity regime, \ie{}, we assume that the scattering can be described in a Minkowski background with quantum-gravitational fluctuations. This allows us to use standard momentum space techniques. While these are limitations that eventually have to be lifted, we expect that such a regime ought to emerge at least approximately in quantum gravity since Minkowski space describes parts of our Universe well.

Our computation is based on the existence of an interacting -- \ie{}~asymptotically safe -- \ac{RG} fixed point. This is necessary to formulate quantum gravity as a fundamental \ac{QFT}. The existence of such a fixed point is, by now, established beyond reasonable doubt, see~\cite{Reuter:2019byg, Pereira:2019dbn, Eichhorn:2020mte, Bonanno:2020bil, Pawlowski:2020qer, Knorr:2022dsx, Eichhorn:2022gku, Morris:2022btf, Martini:2022sll, Wetterich:2022ncl, Platania:2023srt, Saueressig:2023irs, Pawlowski:2023gym, Bonanno:2024xne} for an overview over the results. Most computations that find this fixed point are based on the \ac{FRG}~\cite{Wetterich:1992yh, Morris:1993qb, Ellwanger:1993mw, Gies:2006wv, Dupuis:2020fhh}, where an artificial \ac{IR} cutoff scale is introduced in the path integral. The cutoff gaps modes with momenta smaller than the cutoff scale, while it leaves modes with momenta larger than the cutoff unaltered, resulting in a Wilsonian shell-by-shell integration of fluctuations. In the limit of vanishing cutoff, the standard quantum effective action is recovered. However, recent results~\cite{Basile:2021krr, Knorr:2022ilz, Baldazzi:2023pep, Knorr:2024yiu, Eichhorn:2024wba, DelPorro:2025fiu} provided indications that this limit is more subtle than expected. In particular, some unexpected and unphysical \ac{IR} divergences have been discovered that obstruct the vanishing-cutoff limit. In this paper, we uncover the source of this issue, and we show that in a careful treatment, the effective action obtained from the \ac{FRG} is well-defined and unambiguous.

We learn a handful of important lessons which we expect will inform and shape future investigations in Asymptotic Safety. Our main result is that to avoid the above-mentioned unphysical \ac{IR} divergences, resolving functional momentum dependence (or alternatively, curvature/field dependence) beyond a Taylor expansion is strictly necessary in massless theories. Derivative expansions or simplification techniques like \ac{RG} improvement give quantitatively (in the former case) or even qualitatively (in the latter case) incorrect results. Even in theories with massive degrees of freedom, some of these problems persist in the purely gravitational sector and for classically marginal couplings, whereas they most likely become entirely irrelevant for all practical purposes for all other couplings.

This paper is structured as follows. In \Cref{sec:amp}, we discuss the basics of the scalar scattering amplitude that we are investigating, as well as some relevant approximations. \Cref{sec:AS} covers the basics of the Asymptotic Safety scenario, together with a short overview of the \ac{FRG}, and our precise setup. \Cref{sec:results} contains our results based on resolving the momentum dependence of the massless theory, and a comparison with the derivative expansion as well as \ac{RG} improvement. We also speculate on an effective realisation of the no-global-symmetries conjecture and the potential arisal of a species scale in Asymptotic Safety. In \Cref{sec:results_massive}, we briefly discuss the changes of our results when mass terms for the scalar fields are introduced. In \Cref{sec:consequences}, we discuss how our findings extend to more realistic theories. Lastly, in \Cref{sec:conclusion} we conclude and provide an outlook.

\section{Scalar scattering amplitudes}\label{sec:amp}

In this section, we set the stage and discuss the concrete amplitude that we want to compute in general terms. Specifically, we are interested in two shift-symmetric scalar fields $\phi$ and $\chi$ that are coupled to gravity. Before going into the details, let us briefly discuss the status of symmetries in quantum gravity, as it is relevant to motivate why we are interested in shift-symmetric scalars. There is a widely-believed conjecture that quantum gravity forbids the existence of global symmetries. More precisely, any would-be global symmetry ought to be gauged, or experience Planck-suppressed violations induced by quantum gravity. This conjecture is based on semi-classical black hole arguments and holds in string theory~\cite{Banks:1988yz, Giddings:1987cg, Lee:1988ge, Abbott:1989jw, Coleman:1989zu, Kamionkowski:1992mf, Holman:1992us, Kallosh:1995hi, Banks:2010zn}. It also motivates many of the swampland conjectures~\cite{Vafa:2005ui, Ooguri:2006in, Brennan:2017rbf, Palti:2019pca, vanBeest:2021lhn, Grana:2021zvf, Agmon:2022thq}. Nevertheless, there are examples in which the conjecture does not hold~\cite{Harlow:2020bee} (see, however, \cite{Geng:2025gns}). In Asymptotic Safety all results to date~\cite{Eichhorn:2011pc, Eichhorn:2012va, Labus:2015ska, Percacci:2015wwa, Eichhorn:2017eht, Eichhorn:2020sbo, Ali:2020znq, deBrito:2021pyi, Eichhorn:2021qet, Laporte:2021kyp, Eichhorn:2025ilu, Assant:2025gto} (see also~\cite{Eichhorn:2022gku} for an overview) point towards that global symmetries, including shift symmetry, are not broken by the \ac{RG} flow if the regularisation does not break them artificially. Nevertheless, these results are based on approximations, in particular they are derived in Euclidean signature. It is conceivable that a Lorentzian treatment that properly includes black hole configurations could yield different results. Another option is that Asymptotic Safety gives rise to non-standard black hole thermodynamics~\cite{Basile:2025zjc}, thereby avoiding the semi-classical arguments. Lastly, there could also be a dynamical suppression of black hole configurations in the path integral~\cite{Borissova:2020knn, Borissova:2024hkc}.

With these potential caveats in mind, going forward we assume that shift symmetry is indeed preserved by the \ac{RG} flow, so that we can self-consistently restrict ourselves to the shift-symmetric subspace of the theory. We assume that we can describe the scattering within \ac{QFT} at all energy scales and in a flat Minkowski background. It is useful to describe the most general scattering amplitude in such a setting via the effective action, which can be parameterised as~\cite{Draper:2020knh}
\begin{equation}\label{eq:general_EA}
\begin{aligned}
    \Gamma = \int \text{d}^4x \, \sqrt{-g} \Bigg[ &\frac{-R}{16\pi G_N} + R F_R(\Box)R + R_{\mu\nu} F_{Ric}(\Box) R^{\mu\nu} + \frac{1}{2} \phi \Box Z_\phi(\Box)\phi + \frac{1}{2} \chi \Box Z_\chi(\Box)\chi \\
    &+ F_{R\phi\phi}(\Box_1,\Box_2,\Box_3) R \phi \phi + F_{R\chi\chi}(\Box_1,\Box_2,\Box_3) R \chi \chi \\
    &+ F_{Ric\phi\phi}(\Box_1,\Box_2,\Box_3) R^{\mu\nu} D_\mu\phi D_\nu\phi + F_{Ric\chi\chi}(\Box_1,\Box_2,\Box_3) R^{\mu\nu} D_\mu\chi D_\nu\chi \\
    &- F_{\phi\phi\chi\chi}(D_1\cdot D_2,D_1\cdot D_3, D_1\cdot D_4, D_2 \cdot D_3, D_2 \cdot D_4, D_3\cdot D_4)\phi\phi\chi\chi \Bigg] \, .
\end{aligned}
\end{equation}
Here we are using the mostly minus convention for the metric, and have set the cosmological constant to zero. The form factors $F_R$, $F_{Ric}$, $Z_\phi$, $Z_\chi$, $F_{R\phi\phi}$, $F_{R\chi\chi}$, $F_{Ric\phi\phi}$, $F_{Ric\chi\chi}$, $F_{\phi\phi\chi\chi}$ are functions that depend on scalar combinations of derivatives. These can be d'Alembertians, $\Box=-D^2$, where $D$ is the covariant derivative, or contractions of covariant derivatives acting on different fields, $D_i \cdot D_j$. The subscript $i$ in either case indicates that the corresponding operator acts on the $i$-th field that comes after the form factor. Shift symmetry is implemented in the above action by requiring that $F_{R\phi\phi}$, $F_{R\chi\chi}$ and $F_{\phi\phi\chi\chi}$ vanish in certain limits. For example, the form factors $F_{R\phi\phi}$, $F_{R\chi\chi}$ should be proportional to $D_2\cdot D_3 \equiv (\Box_1 - \Box_2 - \Box_3)/2$. We shall assume that the form factors do not introduce any additional poles in the propagators.

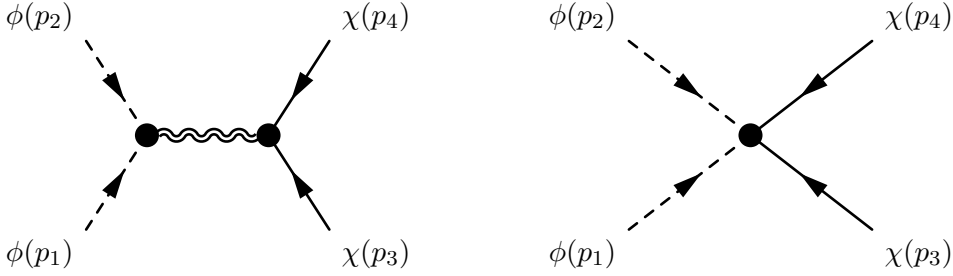
\begin{figure}
\centering
\unitlength = 1mm
\begin{fmffile}{mediated}
\begin{fmfgraph*}(40,25)
\fmfleft{i1,i2}
\fmfright{o1,o2}
\fmflabel{$\phi(p_1)$}{i1}
\fmflabel{$\phi(p_2)$}{i2}
\fmflabel{$\chi(p_3)$}{o1}
\fmflabel{$\chi(p_4)$}{o2}
\fmfv{decor.shape=circle,decor.filled=full,
decor.size=4thick}{v1,v2}
\fmf{scalar}{i1,v1}
\fmf{scalar}{i2,v1}
\fmf{fermion}{o1,v2}
\fmf{fermion}{o2,v2}
\fmf{dbl_wiggly}{v1,v2}
\end{fmfgraph*}
\end{fmffile}
\hspace{2.5cm}
\begin{fmffile}{contactterm}
\begin{fmfgraph*}(40,25)
\fmfleft{i1,i2}
\fmfright{o1,o2}
\fmflabel{$\phi(p_1)$}{i1}
\fmflabel{$\phi(p_2)$}{i2}
\fmflabel{$\chi(p_3)$}{o1}
\fmflabel{$\chi(p_4)$}{o2}
\fmfv{decor.shape=circle,decor.filled=full,
decor.size=4thick}{v1}
\fmf{scalar}{i1,v1}
\fmf{scalar}{i2,v1}
\fmf{fermion}{o1,v1}
\fmf{fermion}{o2,v1}
\end{fmfgraph*}
\end{fmffile}

\vspace{0.5cm}
\caption{\label{fig:amp}Diagrams that contribute to the $\mans$-channel process $\phi(p_1)\phi(p_2)\to\chi(p_3)\chi(p_4)$: gravity-mediated diagram (left) and contact term (right). The wavy double line in the left diagram indicates a fully dressed graviton propagator, and the dotted vertices are fully dressed as well. All momenta are ingoing, as indicated by the arrows.}
\end{figure}

We are interested in the $\mans$-channel process $\phi(p_1)\phi(p_2)\to\chi(p_3)\chi(p_4)$. Since we are using the effective action, only tree-level diagrams have to be evaluated to compute the full scattering amplitude. There are two contributions: a mediated diagram, and a contact term, see \Cref{fig:amp}. The complete expression for the amplitude has been derived in~\cite{Draper:2020knh}. In light of the approximations that we make later on, we only report a simplified amplitude.

For the form factors relevant for the gravity-mediated diagram, we approximate
\begin{equation}
\begin{aligned}
    Z_\phi = Z_\chi &\approx 1 \, , \\
    F_R = F_{Ric} &\approx 0 \, , \\
    F_{R\phi\phi} = F_{R\chi\chi} = F_{Ric\phi\phi} = F_{Ric\chi\chi} &\approx0 \, .
\end{aligned}
\end{equation}
Approximating the wave function renormalisations of the scalars, $Z_\phi$ and $Z_\chi$, by unity has been shown to be an excellent approximation in momentum-dependent computations~\cite{Meibohm:2015twa, Knorr:2019atm, Kher:2025rve}, see also \eg{}~\cite{Dona:2013qba, Dona:2015tnf, Eichhorn:2016esv, Eichhorn:2017eht, Eichhorn:2017sok, Eichhorn:2017als, Eichhorn:2020sbo, deBrito:2021pyi, Laporte:2021kyp, Pastor-Gutierrez:2022nki, deBrito:2025nog} for works computing a constant anomalous dimension, and~\cite{Henz:2016aoh} for a field-dependent wave function. The gravitational form factors $F_R$, $F_{Ric}$ have been computed in various approximations~\cite{Christiansen:2012rx, Codello:2013fpa, Christiansen:2014raa, Bosma:2019aiu, Bonanno:2021squ, Knorr:2021niv, Fehre:2021eob, Knorr:2023usb, Pastor-Gutierrez:2024sbt, Kher:2025rve, Pawlowski:2025etp} and display a non-trivial behaviour which we however neglect to be able to treat the system analytically.\footnote{In our simple approximation, the inclusion of these form factors can largely be compensated by a different regulator choice, see the final flow equation \eqref{eq:LO_flow} below. The only clear exception is the large-momentum behaviour of the propagator, which will be modified under the inclusion of these form factors, in turn adding an anomalous dimension to the form factor that we resolve. The same difference in scaling will also affect the mediated diagram of the scattering amplitude (left panel of \Cref{fig:amp}), so that we expect the overall picture to not change qualitatively.} By contrast, the non-minimal interactions between gravity and shift-symmetric scalar fields have mostly been computed in local approximations~\cite{Eichhorn:2017sok, Laporte:2021kyp, Knorr:2022ilz, Ohta:2025xxo, deBrito:2025nog}, see however~\cite{Chiesa:2026tlz} for recent work on its momentum dependence. See also \eg{}~\cite{Narain:2009fy, Henz:2013oxa, Percacci:2015wwa, Labus:2015ska, Henz:2016aoh, Pawlowski:2018ixd, Wetterich:2019zdo, Eichhorn:2020sbo, Sen:2021ffc, Maitiniyazi:2025pou} for non-minimal interactions between gravity and scalars without shift symmetry.

For the contact term, we make the approximation
\begin{equation}\label{eq:4ptfuncapprox}
    F_{\phi\phi\chi\chi}(\dots)\phi\phi\chi\chi \approx \frac{1}{4} \Big[ (D_\mu\phi)(D_\nu\phi) \Big] \hat X^{\mu\nu\rho\sigma} \Big[ (D_\rho\chi)(D_\sigma\chi) \Big] \, .
\end{equation}
The operator $\hat X$ is spanned by combining metrics, covariant derivatives, and form factors. One admissible expansion into a basis is given by
\begin{equation}\label{eq:FF_in_basis}
\begin{aligned}
    \hat X^{\mu\nu\rho\sigma} &= F_1(\Box) \mathbbm 1^{\mu\nu\rho\sigma} + \frac{1}{4} F_2(\Box) g^{\mu\nu} g^{\rho\sigma} + \frac{1}{2} \left( g^{\mu\nu} F_3(\Box) D^\rho D^\sigma + D^\mu D^\nu F_3(\Box) g^{\rho\sigma} \right) \\
    & + \frac{1}{4} \left( D^\mu g^{\nu\rho} F_4(\Box) D^\sigma + D^\nu g^{\mu\rho} F_4(\Box) D^\sigma + D^\mu g^{\nu\sigma} F_4(\Box) D^\rho + D^\nu g^{\mu\sigma} F_4(\Box) D^\rho \right) \\
    & + D^\mu D^\nu F_5(\Box) D^\rho D^\sigma \, .
\end{aligned}
\end{equation}
Here, we have introduced the identity on symmetric rank-two tensors via
\begin{equation}
    \mathbbm 1^{\mu\nu\rho\sigma} = \frac{1}{2} \left( g^{\mu\rho} g^{\nu\sigma} + g^{\mu\sigma} g^{\nu\rho} \right) \, .
\end{equation}
Different choices of basis and operator ordering only result in differences of the order $\sim R\phi^2\chi^2$, which are irrelevant for the scattering amplitude in Minkowski space. The approximation \eqref{eq:4ptfuncapprox} can be motivated by the observation that if we take free scalar fields that are only minimally coupled to gravity but not to each other, and compute the leading order quantum fluctuations, the latter will be proportional to the respective kinetic terms. In particular, in an Asymptotic Safety setting, such correlation functions that do not break the symmetry of the kinetic term are necessarily non-vanishing at a fixed point~\cite{Eichhorn:2012va}. We will see explicitly that this is realised in our computation below. Local approximations of \eqref{eq:4ptfuncapprox} in the context of gravity have been considered before in \cite{deBrito:2021pyi, Laporte:2021kyp, Knorr:2022ilz, Eichhorn:2022ngh, Ohta:2025xxo}.

We are now in a position to write down the scattering amplitude. The gravity-mediated diagram is simply that known from General Relativity,
\begin{equation}\label{eq:Amp_med}
    \mathcal A_\text{med} = 8 \pi G_N \frac{\mant \manu}{\mans} \, .
\end{equation}
The contact term gives
\begin{equation}
    \mathcal A_\text{cont} = -\frac{1}{16} \bigg[ 2(\mant^2+\manu^2) F_1(\mans) + \mans^2 \Big( F_2(\mans) + 2\mans F_3(\mans) + \mans F_4(\mans) + \mans^2 F_5(\mans) \Big) \bigg] \, .
\end{equation}
Here, we introduced the standard Mandelstam variables\footnote{Note that there are no minus signs in $\mant$ and $\manu$ because of our convention that all momenta are ingoing.}
\begin{equation}
    \mans = (p_1+p_2)^2 \, , \qquad \mant = (p_1+p_3)^2 \, , \qquad \manu = (p_1+p_4)^2 \, .
\end{equation}

To analyse the amplitude further, it is instructive to decompose it into partial wave amplitudes. For this, we first go into the centre-of-mass frame, so that
\begin{equation}
    \mant = -\frac{\mans}{2}(1+x) \, , \qquad \manu = -\frac{\mans}{2}(1-x) \, ,
\end{equation}
where $x=\cos\theta$ is the cosine of the scattering angle between the three-momenta $\vec{p}_1$ and $\vec{p}_3$, $\vec p_1 \cdot \vec p_3 = \sqrt{\vec{p}_1^{\,2}} \sqrt{\vec{p}_3^{\,2}} \,x$. In this frame, the $j$-th partial wave amplitude is defined via
\begin{equation}
    a_j(\mans) = \frac{1}{32\pi} \int_{-1}^1 \text{d}x \, P_j(x) \, \mathcal A(\mans,x) \, .
\end{equation}
The $P_j$ is the $j$-th Legendre polynomial. For our amplitude, only two partial wave amplitudes are non-zero:
\begin{align}
    a_0(\mans) &= \frac{G_N \mans}{12} - \frac{\mans^2}{768\pi} \Big( 4 F_1(\mans) + 3 (F_2(\mans) + 2 \mans F_3(\mans) + \mans F_4(\mans) + \mans^2 F_5(\mans)) \Big) \, , \label{eq:a0} \\
    a_2(\mans) &= -\frac{G_N \mans}{60} - \frac{\mans^2}{1920\pi} F_1(\mans) \, . \label{eq:a2}
\end{align}

In non-gravitational theories, scattering amplitudes need to fulfil a number of properties for the theory to be healthy, including unitarity and causality. A priori, it is not clear that all of them also have to apply to gravitational theories in a straightforward way. In particular, even the very definition of asymptotic states in a curved spacetime is challenging if not impossible, depending on the specific asymptotic structure of spacetime under investigation. For our discussion, we will assume that standard notions like unitarity and causality have to hold, at least in our context of scattering in a flat background.

More specifically, unitarity requires that the partial wave amplitudes must be bounded by unity. We can see that already the tree-level (\ie{}, with all $F_i=0$) gravitational partial wave amplitudes \eqref{eq:a0} and \eqref{eq:a2} violate this at roughly Planckian energies. If the requirement of boundedness carries over to our simplified setup, this would in particular require that the $F_i$ satisfy
\begin{equation}
\begin{aligned}
    F_1(\mans) &\sim -\frac{32\pi G_N}{\mans} \, , \\
    F_2(\mans) + 2\mans F_3(\mans) + \mans F_4(\mans) + \mans^2 F_5(\mans) &\sim \frac{64\pi G_N}{\mans} \, ,
\end{aligned}
\end{equation}
as $\mans\to\infty$, and would thus constitute a necessary requirement to have \emph{physical Asymptotic Safety}, that is at the level of scattering amplitudes~\cite{Weinberg:1980gg, Anber:2010uj, Anber:2011ut, Donoghue:2019clr, Draper:2020bop, Bonanno:2020bil, Donoghue:2024uay}. One of the goals for the remainder of this paper is to compute the functions $F_i$ and check this requirement. More specifically, we will answer whether the existence of an \ac{RG} fixed point with respect to a fiducial \ac{IR} scale $k$ alone implies Asymptotic Safety at the level of scattering amplitudes.

\section{Asymptotic Safety}\label{sec:AS}

In this section, we give an overview of the idea behind Asymptotic Safety, and its technical implementation via the \ac{FRG}. Subsequently, we provide the technical setup that we use to compute the scattering amplitude presented in the previous section.

\subsection{Asymptotic Safety: quantum gravity through scale invariance}

General Relativity, treated as a perturbative \ac{QFT}, is not renormalisable. More specifically, at each loop order in perturbation theory, divergences arise which are not of the form of the original action, as can be seen \eg{}~from the two-loop counterterm~\cite{Goroff:1985sz, Goroff:1985th, VANDEVEN}. This spoils predicitivity as infinitely many free parameters have to be introduced. As an \ac{EFT}, perturbative quantum gravity works extremely well, since the natural cutoff -- the Planck mass -- is extremely large. See \eg{}~\cite{Donoghue:2012zc, Basile:2024oms} for pedagogical introductions to perturbative quantum gravity and its \ac{EFT}.

Another instance where perturbative quantum gravity runs into trouble is the case of scattering amplitudes. As we discussed in the previous section, the partial wave amplitudes of the tree-level gravity-mediated scalar scattering, \eqref{eq:a0} and \eqref{eq:a2}, exceed unity roughly at the Planck scale, indicating a violation of unitarity. The original idea of Asymptotic Safety~\cite{Weinberg:1980gg} is based on the following observation: what if the \ac{RG} running of couplings is such that all relevant scattering amplitudes are bounded? In this case, unitarity could be maintained. The statement that the amplitude is bounded simply means that \emph{dimensionless coupling constants}, that is coupling constants measured in some physical units like energy, \emph{need to go to finite values at high energies}. In other words, the dimensionless quantities have to settle at an \ac{RG} fixed point in the \ac{UV}. This is what we call \emph{physical Asymptotic Safety} for the purpose of this paper.

While physical Asymptotic Safety is certainly necessary for a healthy theory, requiring a fixed point as of itself does not solve the issue of predictivity. After all, we could simply arrange arbitrary form factors in the example of the previous section such that the partial wave amplitudes are bounded.\footnote{That this can actually be arranged when crossing symmetry is taken into account is more subtle, a proof of concept was presented in~\cite{Draper:2020bop}.} Rather, almost all of the newly introduced couplings need to be predictions of the theory. This is the case when they are \ac{RG}-irrelevant, that is, their critical exponents are negative. In summary, we shall define a theory to be physically asymptotically safe if it has a \ac{UV} fixed point with finitely many relevant directions (\ie{}~free parameters), and if all of its scattering amplitudes fulfil unitarity bounds.

A practical investigation of Asymptotic Safety needs a non-perturbative formulation of the \ac{RG}, which we discuss next.

\subsection{Functional renormalisation group}

A useful tool to compute scattering amplitudes is the effective action. Since it contains all quantum fluctuations, tree-level amplitudes contain the complete information about the scattering. While, in principle, the effective action can be computed from the path integral, a more practical approach is based on the Wilsonian idea of integrating out modes momentum-shell by momentum-shell. A specific implementation introduces an interpolating effective action, $\Gamma_k$, which has already taken into account field modes with momenta $p^2\gtrsim k^2$, where $k$ is some \ac{IR} cutoff scale. In the limit $k\to0$, by definition all modes have been integrated out, and the standard effective action $\Gamma$ is obtained. This way of computing the effective action is called the \ac{FRG}.

The $k$-dependence of $\Gamma_k$ can be computed from the Wetterich equation~\cite{Wetterich:1992yh, Morris:1993qb, Ellwanger:1993mw}
\begin{equation}\label{eq:Wetterichequation}
    k \partial_k \Gamma_k = \frac{1}{2} \text{STr} \left[ \left( \Gamma_k^{(2)} + \regulatoroperator_k \right)^{-1} k \partial_k \regulatoroperator_k \right] \, .
\end{equation}
In this, $\Gamma_k^{(2)}$ denotes the second functional derivative of $\Gamma_k$ with respect to the fields, $\regulatoroperator_k$ is a regulator that implements the Wilsonian idea, and STr is the super trace, that is a functional trace (summing/integrating over the eigenvalues of the operator in brackets), an index trace, and a minus sign for fermions. Oftentimes, one introduces an \ac{RG} ``time'' $t=\ln k/k_0$, where $k_0$ is an arbitrary reference scale, and writes an overdot for a $t$-derivative.\footnote{The \ac{RG} time $t$ is not to be confused with the second Mandelstam variable $\mant$. In everything that follows from here on, $t$ always indicates the \ac{RG} time.} For introductions to the \ac{FRG}, see \eg{}~\cite{Percacci:2017fkn, Reuter:2019byg, Dupuis:2020fhh}. The first application of the Wetterich equation to quantum gravity is due to Reuter~\cite{Reuter:1996cp}, for an overview of the state of the art, see~\cite{Knorr:2022dsx, Eichhorn:2022gku, Morris:2022btf, Martini:2022sll, Wetterich:2022ncl, Platania:2023srt, Saueressig:2023irs, Pawlowski:2023gym, Bonanno:2024xne}.

While \eqref{eq:Wetterichequation} is formally exact, in practice approximations are necessary to obtain results. The standard procedure to investigate Asymptotic Safety with the \ac{FRG} is thus based on choosing an approximation for $\Gamma_k$ spanned by a number of $k$-dependent couplings, computing beta functions by evaluating \eqref{eq:Wetterichequation}, searching for fixed points for $k\to\infty$, analysing the critical exponents, and finally computing the \ac{IR} values of the couplings as $k\to0$, \ie{}~their physical values in the effective action $\Gamma$. The number of relevant directions dictates the dimension of the manifold of possible outcomes of this procedure, \ie{}~the asymptotically safe landscape. The computation of the full landscape has received a lot of interest lately~\cite{Basile:2021krr, Knorr:2022ilz, Knorr:2024yiu, Saueressig:2024ojx, Basile:2025zjc, DelPorro:2025fiu, DelPorro:2025wts}.

A common approximation scheme is the derivative expansion, which proposes to expand the effective action in powers of derivatives. For our concrete example, this would entail an expansion of the form factors in a Taylor series. This means that we would make the ansatz
\begin{equation}
    F_k^{\text{DE}}(P^2) = \sum_{n\geq0} C_{n,k} P^{2n} \, ,
\end{equation}
where the couplings $C_{n,k}$ carry the $k$-dependence. Its benefit is that instead of solving differential or integro-differential equations for form factors as functions of both $k$ and $P$, the task is reduced to solving first-order differential equations in $k$ for the couplings $C_{n,k}$, which is significantly simpler. A key requirement for the scheme to work is that the above series converges to the full result. This will be challenged by our results below.

For our exposition, it is important to highlight two aspects in dealing with the Wetterich equation. First of all, the above formulation uses a Euclidean-signature metric. This is because cutting modes with high/low energy is involved with Lorentzian momenta -- one can have arbitrarily high energies/frequencies while keeping the norm of the momentum small. Recently, formulations in Lorentzian signature have been successfully implemented~\cite{Manrique:2011jc, Bonanno:2021squ, Fehre:2021eob, Banerjee:2022xvi, Saueressig:2023tfy, DAngelo:2023wje, DAngelo:2023tis, Ferrero:2024rvi, Banerjee:2024tap, Thiemann:2024vjx, Pastor-Gutierrez:2024sbt, Saueressig:2025ypi, Kher:2025rve, DAngelo:2025yoy, Pawlowski:2025etp}, but the bulk of the literature on Asymptotic Safety is still based on Euclidean computations. We also use the Euclidean formulation here, but since we have full analytical control, we can perform the Wick rotation back to Lorentzian signature at the very end.

Second, the role of the momentum scale $k$ is really an auxiliary one, with a priori no physical meaning~\cite{Knorr:2019atm, Bonanno:2020bil}. It is not to be confused with a physical momentum, meaning a derivative in the effective action, or a momentum in a scattering process~\cite{Buccio:2023lzo, Buccio:2024hys, Buccio:2025tci}. This entails that some couplings can have a non-trivial \ac{RG} running with $k$ but not with $p$ -- this includes in particular Newton's constant and the cosmological constant. This distinction gave rise to some confusion in the literature~\cite{Anber:2011ut, Donoghue:2019clr, Donoghue:2024uay}. Nevertheless, in \emph{some} situations, the $k$-running mimics the $p$-running~\cite{Bonanno:2020bil}, \eg{}~when there is a decoupling of modes below a mass threshold~\cite{Appelquist:1974tg, Reuter:2003ca, Borissova:2022mgd}, but this can in general only be verified a posteriori.

A simplification technique that assumes that $k$- and $p$-dependence can be identified is \ac{RG} improvement~\cite{Dittrich:1985yb}: instead of resolving a finite number of coefficients $C_{n,k}$ of the derivative expansion to at least approximately resolve the $P$-dependence, \ac{RG} improvement typically only takes the leading order term $C_{0,k}$ and identifies the \ac{FRG} scale $k$ with the momentum scale $P$ (or another physical scale like curvature). This process is in general non-unique, and by construction cannot describe multi-scale physics (which generally includes scattering processes). Other issues include the potential breaking of diffeomorphism symmetry or Bianchi identities. Nevertheless, this procedure has been used widely to derive Asymptotic-Safety-inspired phenomenology for black holes and cosmology -- see~\cite{Platania:2023srt, Bonanno:2024xne} for an overview, while its limitations have been discussed \eg{}~in~\cite{Donoghue:2019clr, Bonanno:2020bil, Held:2021vwd, Borissova:2022mgd}. The key difference to the derivative expansion is that \ac{RG} improvement is not a systematic expansion scheme -- it is not expected to converge to the true result, but rather to capture the qualitative picture. This expectation will be challenged by our results.

In our system, we will be able to track the full $k$- and $p$-dependence of the form factor to evaluate the quality of such a scale identification. As a matter of fact, this will yield insights into which approximations can or cannot converge to the true effective action.

\subsection{Setup}

To investigate scalar scattering amplitudes, we make the following approximation for the effective action:
\begin{equation}
\begin{aligned}
    \Gamma_k^E = \int \text{d}^4x \, \sqrt{g} \, \bigg[  -\frac{1}{16\pi G_{N,k}} R + \frac{1}{2} (D_\mu\phi)&(D^\mu\phi) + \frac{1}{2} (D_\mu\chi)(D^\mu\chi) \\
    &+ \frac{1}{4} (D_\mu\phi)(D_\nu\phi) \hat X_k^{E\mu\nu\rho\sigma} (D_\rho\chi)(D_\sigma\chi) \bigg] \, .
\end{aligned}
\end{equation}
With the superscript $E$ we indicate that we are in Euclidean signature. In analogy with \eqref{eq:FF_in_basis}, the operator $\hat X_k$ is now spanned by $k$-dependent Euclidean form factors $F_{i,k}^E(\Delta)$, which depend on the Laplacian $\Delta=-D^2$. Note that compared with the Lorentzian version \eqref{eq:general_EA}, the Wick rotation consists loosely speaking in an overall minus sign, and an extra minus sign for every two derivatives.\footnote{A more careful treatment of Wick rotation in quantum gravity is actually involved, and might not exist even in highly symmetric spaces~\cite{Baldazzi:2018mtl}. Since our setup lives in a flat background, we do not expect conceptual issues to arise (beyond any that exist for non-gravitational theories anyway), but a more complete treatment might face issues.} We also want to draw attention to the explicit relative minus sign in front of the form factor, which has been introduced to avoid a minus sign in the Wick rotation.

\begin{figure}
\centering
\unitlength = 1mm
\begin{fmffile}{seagull}
\begin{fmfgraph*}(40,25)
\fmfleft{i1,i2}
\fmfright{o1,o2}
\fmflabel{$\phi(p_1)$}{i1}
\fmflabel{$\phi(p_2)$}{i2}
\fmflabel{$\chi(p_3)$}{o1}
\fmflabel{$\chi(p_4)$}{o2}
\fmfv{decor.shape=circle,decor.filled=full,decor.size=4thick}{v1,v3}
\fmf{scalar}{i1,v1}
\fmf{scalar}{i2,v1}
\fmf{fermion}{o1,v3}
\fmf{fermion}{o2,v3}
\fmf{dbl_wiggly,left,tension=0.3}{v1,v3,v1}
\end{fmfgraph*}
\end{fmffile}

\vspace{0.5cm}
\caption{\label{fig:flow}Double seagull diagram that we have to evaluate for the flow of the form factors $F_{i,k}^E(P^2)$. The wavy double lines represent graviton propagators, whereas the full and dashed lines correspond to the scalar fields. A regulator insertion $k\partial_k \regulatoroperator_k$ is understood on one of the propagator lines.}
\end{figure}
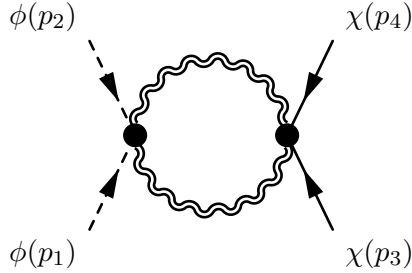

We are interested in computing the \ac{RG} flow of the form factors $F_{i,k}^E$. For this, we implement several simplifications to make the computation tractable. First, we neglect any quantum fluctuations of the scalar fields, \ie{}, they only appear as external lines. Second, we neglect the self-feedback of the form factors into their own flow equation. These two approximations imply that the only diagram that contributes to the flow of $F_{i,k}^E$ is a double seagull diagram, see \Cref{fig:flow}, which justifies the simplified momentum dependence.

To be able to discuss fixed points, we first have to introduce dimensionless quantities $g_k, f_{i,k}$ via
\begin{equation}
    G_{N,k} = g_k k^{-2} \, , \qquad F_{i,k}^E(P^2) = k^{-4} f_{i,k}^E(P^2/k^2) \, .
\end{equation}
In the following, it will be useful to view $g_k$ and $f_{i,k}^E$ as functions of either $k$ or the \ac{RG} ``time'' $t$, and we make this explicit by writing the corresponding subscript. For example, $g_k$ indicates the dimensionless Newton's constant as a function of $k$, whereas $g_t$ indicates the same function but with the replacement $k=k_0 e^t$.

Additionally, we approximate the \ac{RG} flow of Newton's coupling by a simple flow equation that captures all qualitative features. Our simple approximation reads
\begin{equation}
    k \partial_k g_k = 2g_k \left( 1 - \frac{g_k}{g_\ast} \right) \equiv (2+\eta_{N,k})g_k \, .
\end{equation}
In this, $g_\ast>0$ is the asymptotically safe fixed point value of the dimensionless Newton's coupling $g_k$, and $\eta_{N,k}$ is its anomalous dimension. This equation has the solution
\begin{equation}
    g_k = \frac{G_N k^2}{1 + \frac{G_N k^2}{g_\ast}} \, ,
\end{equation}
which approximates actual solutions in advanced investigations surprisingly well~\cite{Bonanno:2000ep, Fehre:2021eob}. Note that $G_N$ (without the subscript $k$) indicates the physical Newton's coupling at vanishing \ac{IR} cutoff, $G_N = \lim_{k\to0} G_{N,k}$. Finally, we set the cosmological constant to zero and neglect its flow. While the overall setup constitutes a severe simplification, it allows us to understand all aspects of momentum-dependent flows analytically.

We amend the action with a standard harmonic gauge fixing since this leads to the simplest possible momentum dependence. The ensuing Faddeev-Popov ghosts play no role for our considerations. Lastly, we use the regulator
\begin{equation}
    \regulatoroperator_k^{\mu\nu\rho\sigma} = \frac{1}{16\pi G_{N,k}} \left[ \mathbbm 1^{\mu\nu\rho\sigma} - \frac{1}{2} g^{\mu\nu} g^{\rho\sigma} \right] \, R_k(\Delta)  \, .
\end{equation}
For the shape function, we choose
\begin{equation}\label{eq:expshapefunction}
    R_k(z) = \frac{z}{e^{\frac{z}{k^2}}-1} \, ,
\end{equation}
which enables us to perform all integrals analytically.

All tensor algebra was handled with the package \emph{xAct}~\cite{Martin-Garcia:2007bqa,Brizuela:2008ra,Martin-Garcia:2008yei,Martin-Garcia:2008ysv,Nutma:2013zea}.

\section{Results for the massless theory}\label{sec:results}

In this section, we present the results of our analytical, momentum-dependent computation for shift-symmetric scalar fields. We discuss several aspects of the solution, connect to perturbation theory and \ac{EFT}, and evaluate the partial wave amplitudes for the scalar scattering process. We then compare our results to those obtained in a derivative expansion, and show that it falls short of the analytical result in a quantitative way. We also show that \ac{RG} improvement fails qualitatively to represent the physical momentum dependence.

\subsection{Momentum-dependent results}

We now derive the prediction for the form factors $F_{i,k}^E(\Delta)$ within our setup. The reader who is only interested in the result as well as physical consequences can skip to \eqref{eq:FsolLOIR} and the subsequent discussion.

We evaluate the double seagull diagram of \Cref{fig:flow} in a flat background so that we can use momentum space techniques. We choose all momenta as ingoing, and we introduce the momentum transfer $P$ as
\begin{equation}
    P_\mu = p_{1,\mu} + p_{2,\mu} = - p_{3,\mu} - p_{4,\mu} \, .
\end{equation}
With this, we find that
\begin{equation}\label{eq:LO_flow}
    k \partial_k \hat X_k^{E\mu\nu\rho\sigma} = 16 \Pi_\text{TL}^{\mu\nu\rho\sigma} (16\pi G_{N,k})^2 \int \frac{\text{d}^4q}{(2\pi)^4} \frac{\dot R_k(q^2) - \eta_{N,k} R_k(q^2)}{\left( q^2 + R_k(q^2)\right)^2 \left( (P+q)^2 + R_k((P+q)^2) \right)} \, .
\end{equation}
Here, we used the traceless projector
\begin{equation}
    \Pi_\text{TL}^{\mu\nu\rho\sigma} = \mathbbm 1^{\mu\nu\rho\sigma} - \frac{1}{4} g^{\mu\nu} g^{\rho\sigma} \, .
\end{equation}
For our general ansatz \eqref{eq:FF_in_basis} this means that $F_3=F_4=F_5=0$ so that no potentially non-local terms appear, and $F_1=-F_2\equiv F$. To simplify the notation, we thus introduce the single form factor $F^E$ via
\begin{equation}
    \hat X_k^{E\mu\nu\rho\sigma} = \Pi_\text{TL}^{\mu\nu\rho\sigma} F_k^E(P^2) \, .
\end{equation}

To solve the flow equation, we switch to dimensionless quantities. For this, we introduce the dimensionless squared momentum
\begin{equation}
    z = \frac{P^2}{k^2} \, .
\end{equation}
As explained before, we indicate explicitly whether we consider the functions to depend on the dimensionful \ac{RG} scale $k$, or the dimensionless \ac{RG} time $t$. To solve the flow equation, we first treat everything as a function of $t$, so that
\begin{equation}
    \dot f_t^E(z) - 4 f_t^E(z) -2z f_t^{E\prime}(z) = (16\pi g_t)^2 I(z) \, ,
\end{equation}
where we abbreviated the dimensionless integral in \eqref{eq:LO_flow} by $I$, including a prefactor of $16$. For our regulator choice, it reads explicitly
\begin{equation}
    I(z) = \frac{2}{\pi^2} \left[ \frac{1-e^{-\frac{z}{2}}}{z} - \frac{\eta_{N,t}}{4} \left( \frac{1 + 3 e^{-\frac{2z}{3}} - 4 e^{-\frac{z}{2}}}{z} - 2 \left( \Gamma(0,\tfrac{2z}{3}) - \Gamma(0,\tfrac{z}{2}) \right) \right) \right] \, .
\end{equation}
In this expression, $\Gamma(a,b)$ is the incomplete Gamma function.

To improve the clarity of our exposition, in the main text we neglect the term proportional to $\eta_{N,t}$. In the appendix, we show that the qualitative behaviour is unaltered by this approximation, and even the quantitative error is limited.

With this approximation, we thus have to solve
\begin{equation}\label{eq:dimlessflowequationLO}
    \dot f_t^E(z) - 4 f_t^E(z) -2z f_t^{E\prime}(z) = 512 g_t^2 \, \frac{1-e^{-\frac{z}{2}}}{z} \, .
\end{equation}
The general solution to this differential equation can be given in integral form:
\begin{equation}\label{eq:LO_dimless_solution_integral_form}
    f_t^E(z) = \frac{256}{z^2} \int_0^z \text{d}y \, \left( e^{-\frac{y}{2}} - 1 \right) \, g_{t + \frac{1}{2}\ln \frac{z}{y}}^2 + \frac{c(t+\frac{1}{2}\ln z)}{z^2} \, .
\end{equation}
In this, $c$ is a free function. It is uniquely fixed to be zero by requiring regularity at vanishing momenta for all finite $t$ as well as fixed-point scaling when $t\to\infty$.\footnote{This result can also be arrived at by a traditional fixed point analysis. The single integration constant of the fixed point solution is fixed by regularity at $z=0$. A stability analysis then reveals the spectrum of critical exponents to be $\theta\in\{2,-4-2\mathbbm N_0 \}$ by requiring regularity of the eigenfunctions at $z=0$. The fact that the critical exponents are exactly related to the mass dimensions of the included operators is due to the simplicity of our approximation.} Note that this entails that in our setup, there are no other relevant parameters besides Newton's constant. In other words, the whole form factor $f$ corresponds to an irrelevant operator at the fixed point. This is in agreement with earlier local investigations, where a similar fixed point was found~\cite{deBrito:2021pyi, Laporte:2021kyp, Knorr:2022ilz, Eichhorn:2022ngh}.

If we now insert our approximation for the scale dependence of $g_t$, we can perform the remaining integral. To present the result, it is useful to switch back to $k$, so that
\begin{equation}\label{eq:LO_dimless_solution}
\begin{aligned}
    f_k^E(z) = &-\frac{256 g_\ast (G_N k^2)^2}{g_\ast + G_N k^2} \frac{e^{-\frac{z}{2}} - 1}{z} \\
    &- 128 (G_N k^2)^2 e^{\frac{(g_\ast+G_N k^2) z}{2g_\ast}} \left( \Gamma(0,\tfrac{G_N k^2 z}{2g_\ast}) - \Gamma(0,\tfrac{(g_\ast + G_N k^2)z}{2g_\ast}) \right) \, .
\end{aligned}
\end{equation}
We emphasise once again that in this expression (and all other expressions below), $G_N$ is the physical value of Newton's constant, \ie{}, $G_N = G_{N,0}$. In the solution, we see that all factors of the \ac{RG} scale $k$ are measured in units of $G_N$.

\begin{figure}[t]
\centering
\includegraphics[width=.7\linewidth]{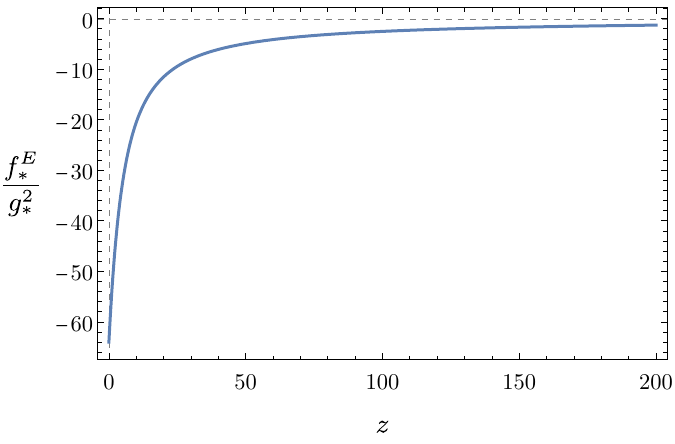} 
\caption{\label{fig:LO_FP}Fixed point solution $f_\ast^E(z)$ in units of the fixed point value of Newton's coupling, as a function of the dimensionless Euclidean squared momentum $z=p^2/k^2$. The function is finite at zero argument, and falls off like $1/z$ for large $z$. It is negative over the entire domain.}
\end{figure}

Let us analyse this solution. First of all, we are reaching the fixed point solution for $k\to\infty$:
\begin{equation}\label{eq:FPLOsol}
    f_\ast^E(z) = \lim_{k\to\infty} f_k^E(z) = - \frac{256g_\ast^2}{z^2} \left( 2 e^{-\frac{z}{2}} - 2 + z \right) \, .
\end{equation}
Note that all factors of the \ac{IR} quantity $G_N$ drop out, as they must. This expression is regular for all $z\geq0$. In particular, expanding in powers of $z$ about zero, we find the everywhere convergent power series
\begin{equation}
    f_\ast^E(z) = -128g_\ast^2 \sum_{n\geq0} \frac{1}{(n+2)!} \left( -\frac{z}{2} \right)^n \, ,
\end{equation}
proving finiteness for small $z$. For large $z$, we have the asymptotic relation
\begin{equation}\label{eq:FPLO_large_z}
    f_\ast^E(z) \sim -\frac{256g_\ast^2}{z} \, , \qquad z \to \infty \, .
\end{equation}
The fixed point solution is displayed in \Cref{fig:LO_FP}.

Next, we investigate the physical limit $k\to0$. For this, we have to go back to the dimensionful form factor, and reinstate the Euclidean squared momentum $P^2 = z k^2$. Only then can we safely take the $k\to0$ limit. We obtain
\begin{equation}\label{eq:FsolLOIR}
    F^E(P^2) = -128 G_N^2 \, e^{\frac{G_N P^2}{2g_\ast}} \, \Gamma(0,\tfrac{G_N P^2}{2g_\ast}) \, .
\end{equation}
This result is remarkable in several aspects, that we discuss now.

\paragraph{Finiteness.}

The limit $k\to0$ can be taken safely, without encountering any divergences or ambiguities. In other words, there is a \emph{unique and finite} prediction for this form factor, given our asymptotically safe fixed point.\footnote{The fact that the prediction is unique is due to the fact that we do not have any additional relevant directions that could introduce free parameters, cf.~the discussion of the homogeneous solution $c$ below \eqref{eq:LO_dimless_solution_integral_form}.} This is a non-trivial statement in the light of recent results~\cite{Basile:2021krr, Knorr:2022ilz, Baldazzi:2023pep, Knorr:2024yiu, Eichhorn:2024wba, DelPorro:2025fiu}, which find divergences in this limit. We explain the origin of such spurious divergences in \Cref{sec:DE}.

\paragraph{The scale of quantum gravity in Asymptotic Safety.}

From the solution \eqref{eq:FsolLOIR}, we can identify the scale
\begin{equation}\label{eq:QGscale}
    \Lambda_\text{EFT}^2 \simeq \frac{2g_\ast}{G_N} \, .
\end{equation}
The obvious interpretation of this scale would be the cutoff scale of the corresponding \ac{EFT}, or equivalently the scale where quantum gravity effects become important. We can actually fix the precise form of the \ac{EFT} cutoff if we compare to a one-loop computation. Evaluating the diagram \Cref{fig:flow} with a simple momentum cutoff $\Lambda_\text{EFT}$ yields the one-loop result\footnote{To evaluate this, we have chosen internal momenta of propagators as $q$ and $P+q$, respectively, and imposed the cutoff on the absolute value of the loop momentum $q$.}
\begin{equation}
    F_\text{EFT}^E(P^2) = 128G_N^2 \left( \ln \frac{P^2}{\Lambda_\text{EFT}^2} - 1 \right) \, .
\end{equation}
We find a logarithmic divergence of the local term, as expected from power counting. Comparing to the small momentum behaviour of the full solution \eqref{eq:FsolLOIR}, we can read off the \ac{EFT} cutoff:
\begin{equation}\label{eq:EFTcutoffLO}
    \Lambda_\text{EFT}^2 =  \frac{2g_\ast}{G_N} e^{-(1+\gamma)} \, .
\end{equation}
In this, $\gamma$ is Euler's gamma constant. Notably, the cutoff is not just the Planck scale $G_N$, but it is scaled by the fixed point value $g_\ast$. In particular, if the fixed point is very non-perturbative, \ie{}~$g_\ast\gg1$, the scale $\Lambda_\text{EFT}$ is much larger than the Planck mass. On the other hand, if we are in a near-perturbative scenario, $g_\ast\ll1$, and the scale is much lower than the Planck mass.\footnote{We note in passing that a priori, fixed point values are not a good indicator for the level of perturbativity, since couplings can be rescaled without changing the physics. Nevertheless, absent arbitrarily large or small rescalings, the size of fixed point couplings dictate the size of critical exponents, which are invariant under rescaling of couplings. It is in this sense that perturbativeness should be understood here: given a fixed normalisation, we expect that the larger $g_\ast$, the more non-canonical the critical exponents, and thus the more non-perturbative the fixed point.} Note that in purely gravitational systems in Asymptotic Safety, fixed point values are typically $\mathcal O(1)$, so that the naive cutoff scale $\Lambda_\text{EFT}^2\sim 1/G_N$ is quantitatively correct. However, for some matter content,\footnote{The most advanced approximations indicate that only vector fields anti-screen, whereas scalars and fermions screen the gravitational coupling, see~\cite{Eichhorn:2022gku} for an overview of results.} the fixed point value $g_\ast$ decreases as the number of fields increases, thus significantly reducing the actual cutoff scale. We speculate that this might be a mechanism for how the species scale known from string theory or arguments based on the size of a smallest-possible black hole~\cite{Dvali:2001gx, Veneziano:2001ah, Dvali:2007wp, Cribiori:2022nke, Cribiori:2023ffn, Basile:2023blg, Basile:2024dqq, ValeixoBento:2025bmv} can arise in Asymptotic Safety, see also~\cite{Dona:2013qba}.

\paragraph{Scale transmutation.}

In perturbation theory, the power of Newton's coupling of the leading-order quantum-gravitational contribution to a specific beta function is dictated solely by the mass dimension of the coupling. For example, relative to the Einstein-Hilbert term, the gravitational one-loop counterterm has a single factor of $G_N$, the two-loop counterterm has a factor of $G_N^2$ and so on. Structurally, the effective action would thus scale as
\begin{equation}
    \Gamma \simeq \frac{1}{16\pi G_N} \int \text{d}^4x \, \sqrt{-g} \, \bigg[ R + c_1 G_N R^2 + c_2 G_N^2 R^3 + c_3 G_N^2 R \Box R + \dots \bigg] \, ,
\end{equation}
where the $c_i$ are pure numbers, and we only indicated some of the terms. For our form factor, this power counting argument suggests that every factor of $P^2$ ought to come with a factor of $G_N$. This would then naively suggest that if we were to expand the form factor in powers of the momentum, the beta functions of the $n$-th term would need to scale with a corresponding power of $G_N$. This general observation is to be contrasted with the form of the flow equation \eqref{eq:LO_flow}, where the quantum term is exactly quadratic in $g_k$. This would naively suggest that our result is purely one-loop. In particular, powers of the external momenta (which in perturbation theory would give rise to extra factors of $G_N$, as argued above) do not come with extra factors of $g_k$, and one might be worried that this leads to inconsistencies.

The resolution of this apparent conflict can be seen in the solution \eqref{eq:LO_dimless_solution_integral_form}. Effectively, the \ac{RG} scale $k$ is transmuted to the physical scale $G_N$ as $k\to0$. This is an aspect that is broken in a derivative expansion, which we discuss in more detail in \Cref{sec:DE}. Also, clearly our solution includes all-order effects, see the previous comparison to the actual one-loop result.

\paragraph{Analytic structure.}

\begin{figure}[t]
\centering
\includegraphics[width=.7\linewidth]{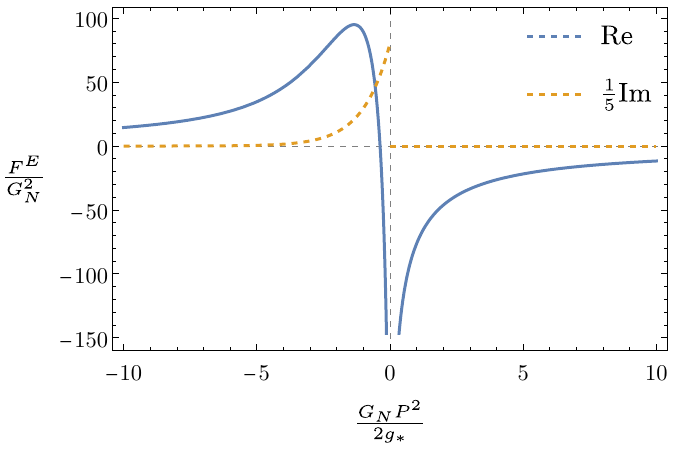} 
\caption{\label{fig:LO_IR}Real and imaginary parts of the Euclidean form factor $F^E$ in Planck units, as a function of Euclidean squared momenta in terms of the quantum gravity scale $G_N/2g_\ast$. At vanishing argument, there is a logarithmic divergence in the real part, see \eqref{eq:LOFFexpanded}.}
\end{figure}

\begin{figure}[t]
\centering
\includegraphics[width=.7\linewidth]{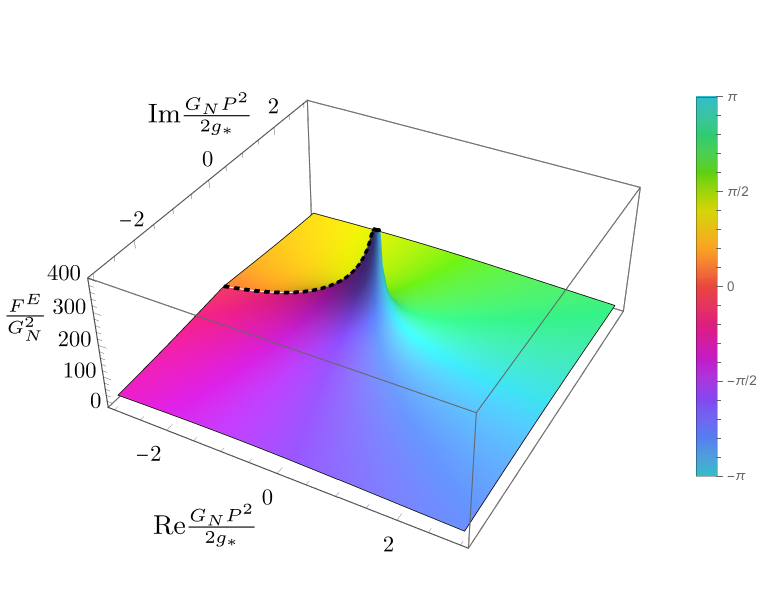} 
\caption{\label{fig:LO_IR_complex}Euclidean form factor $F^E$ in Planck units, as a function of complex Euclidean squared momenta in terms of the quantum gravity scale $G_N/2g_\ast$. The colour gradient indicates the phase of the function, and the dashed line indicates the branch cut.}
\end{figure}

The form factor is analytic everywhere in the complex plane, except for a branch cut on the negative real axis. This is expected as it corresponds to the multi-particle continuum. In particular, along the cut, we have
\begin{equation}\label{eq:branchcut}
    \text{Im} \, F^E(P^2) = 128 \pi G_N^2 \, e^{\frac{G_N P^2}{2g_\ast}} \, , \qquad P^2 \leq 0 \, .
\end{equation}
This allows us to perform the Wick rotation and evaluate the form factor for squared momenta $P^2\in\mathbbm C$.
We show the form factor on the real $P^2$-axis in \Cref{fig:LO_IR}, and in the complex plane in \Cref{fig:LO_IR_complex}.

The form factor has an expansion in powers of the momentum and its logarithm:
\begin{equation}\label{eq:LOFFexpanded}
    F^E(P^2) = 128 G_N^2 \sum_{n\geq0} \frac{1}{n!} \left[ -\psi(n+1) +  \ln \frac{G_N P^2}{2g_\ast} \right] \left( \frac{G_N P^2}{2g_\ast} \right)^n \, .
\end{equation}
In this, $\psi$ is the digamma function, and we can see explicitly how the scale \eqref{eq:QGscale} emerges. We emphasise that the logarithm dominates the constant term for almost all physical scales below the Planck scale. The expansion shows how physical \ac{RG} running should be interpreted: the couplings that multiply polynomials of momentum measured in the (natural) units of the theory have a logarithmic dependence on this momentum. Notably, these polynomial couplings do \emph{not} show a fixed point individually at high energies, as all of them grow logarithmically -- nevertheless, the resummed form factor goes to zero. This is to be contrasted with the usual interpretation of \ac{RG} running in the \ac{FRG}, where couplings generically show power-law running.

For large (complex) momentum, the form factor falls off,
\begin{equation}
    F^E(P^2) \sim -\frac{256 g_\ast G_N}{P^2} \, , \qquad |P^2| \to \infty \, , P^2 \notin \mathbbm R_- \, .
\end{equation}
Intriguingly, the numerical coefficient is exactly the same as the one obtained for the fixed point solution, \eqref{eq:FPLO_large_z}. The only difference is the replacement of one of the fixed point values $g_\ast$ by a dimensionful Newton's constant $G_N$, which is entirely due to accounting for the correct mass dimension. This relation is independent of the choice of regulator shape function in our system. If this pattern persists in more complete treatments, the analysis of scattering amplitudes at large energies would become much easier, since solving fixed point equations is significantly simpler than solving full flows. The pattern might be related to the presence of momentum locality~\cite{Christiansen:2014raa, Christiansen:2015rva, Pawlowski:2020qer, Knorr:2021niv}, which is the property that the flow of a correlation function in units of itself goes to zero as any of its momenta goes to infinity. Intuitively, this would mean that the flow at large momenta is not strong enough to alter the correlation function, leading to a fixed asymptotic behaviour.

\paragraph{Momentum beta function.}

\begin{figure}[t]
\centering
\includegraphics[width=.7\linewidth]{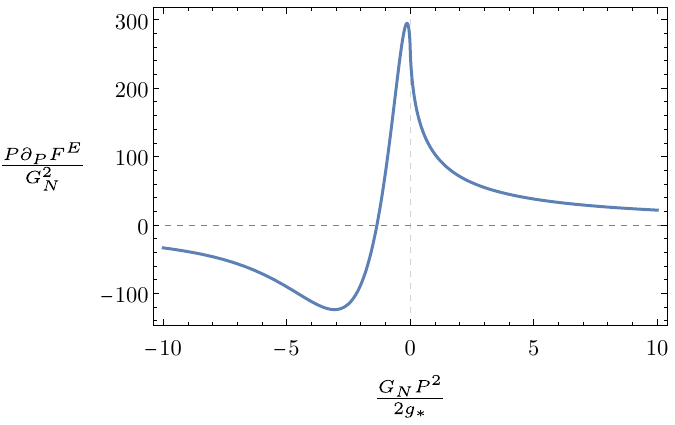} 
\caption{\label{fig:LO_physbeta}Real part of the momentum beta function in Planck units, as a function of Euclidean momenta in terms of the quantum gravity scale. Since we have a completely analytical expression, we can evaluate the form factor also for negative arguments. At vanishing argument, the function is continuous but not differentiable due to the logarithm in the form factor, cf. \eqref{eq:LOFFexpanded} and \eqref{eq:LOmombeta}.}
\end{figure}

With the analytic momentum dependence at hand, we can derive a momentum beta function for the form factor in Planck units:
\begin{equation}\label{eq:LOmombeta}
    P \partial_P F^E(P^2) = 256 G_N^2 \left( 1 - \frac{G_N P^2}{2g_\ast} \, e^{\frac{G_N P^2}{2g_\ast}} \Gamma(0,\tfrac{G_N P^2}{2g_\ast}) \right) = 256 G_N^2 + 2 \frac{G_N P^2}{2g_\ast} \, F^E(P^2) \, .
\end{equation}
We depict its real part in \Cref{fig:LO_physbeta}. We can see that for large (negative or positive) $P^2$, this beta function goes to zero. In this sense, the form factor displays a fixed point at large momentum.

\paragraph{Scalar scattering.}

To discuss the scalar scattering, we have to Wick-rotate our result. With our conventions of a mostly minus signature for the Minkowski metric, and the explicit sign choice for the contact term as in \eqref{eq:general_EA}, we have
\begin{equation}
    F_1(\mans) = -F_2(\mans) = F^E(-\mans) \, , \qquad F_3=F_4=F_5=0 \, .
\end{equation}
The form factors without superscript are those in Lorentzian signature, \ie{}~the physically relevant ones. Since $\mans>0$, we have to access negative squared Euclidean momenta, $P^2\to-\mans$. This entails that
\begin{equation}
    F_1(\mans) = -F_2(\mans) = -128G_N^2 e^{-\frac{G_N \mans}{2g_\ast}} \Gamma(0,-\tfrac{G_N \mans}{2g_\ast}) \, .
\end{equation}
Using our general formula for the partial wave amplitudes, we find at large energies $\mans\to\infty$,\footnote{Since the imaginary part falls off exponentially, it plays no role for our considerations here.}
\begin{align}
    a_0(\mans) &\sim \frac{\pi-4g_\ast}{12\pi} G_N \mans - \frac{2g_\ast^2}{3\pi} \, , \\
    a_2(\mans) &\sim -\frac{\pi+8g_\ast}{60\pi} G_N \mans - \frac{4g_\ast^2}{15\pi} \, .
\end{align}
We conclude that in our simple setup, there is no possible fixed point value $g_\ast$ to bound both partial wave amplitudes. In particular, the spin two partial wave amplitude receives corrections of the ``wrong'' sign since $g_\ast>0$. As a consequence, \emph{the mere existence of a fixed point at large $k$ does not necessarily imply bounded scattering amplitudes, at least in approximations}.

\begin{figure}[t]
\centering
\includegraphics[width=\textwidth]{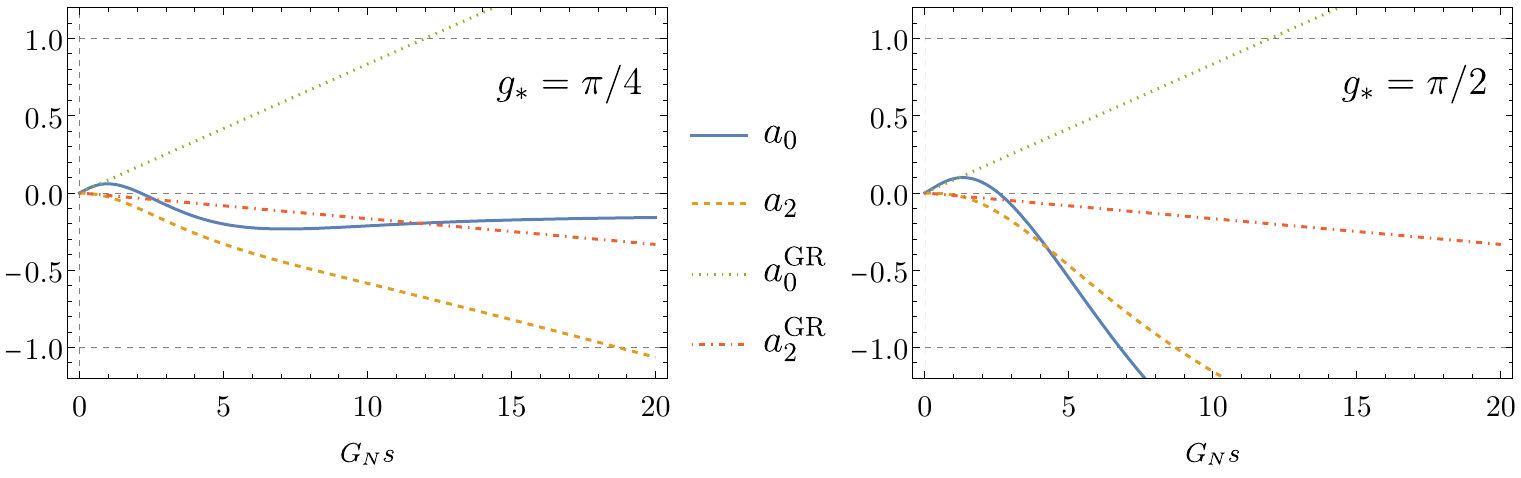} 
\caption{\label{fig:PWA}Comparison of real parts of the partial wave amplitudes with their tree-level counterparts as a function of energy transfer in Planck units, for two choices of the fixed point value $g_\ast$. In the left panel, $g_\ast=\pi/4$ so that $a_0$ is bounded, but $a_2$ is not bounded and exceeds unit strength before the tree-level amplitude does. In the right panel, $g_\ast=\pi/2$, and both $a_0$ and $a_2$ exceed unity before their tree-level counterparts.}
\end{figure}

We show the real parts of the partial wave amplitudes in comparison to their tree-level counterparts in \Cref{fig:PWA}, for two choices of $g_\ast$. In the left panel, we choose $g_\ast=\pi/4$ which bounds $a_0$ for all energies. In the right panel, we choose $g_\ast=\pi/2$, for which both $a_0$ and $a_2$ are unbounded. Both choices of $g_\ast$ are broadly compatible with the literature. We find that for these choices, the partial wave amplitudes violate the unitarity bound at \emph{lower} energies than at tree level -- except of course for $a_0$ and $g_\ast=\pi/4$.

Our approximation is very simple, and a more complete treatment, including the momentum dependence of the propagators and all relevant vertices, has to be carried out. Only then, one can make the final verdict about whether the scattering amplitude is asymptotically safe. Nevertheless, our simple system illustrates -- from start to end -- how such a more complete analysis can be realised.

\paragraph{(No) Global symmetries in Asymptotic Safety?}

Let us assume for a moment that in a more complete treatment, the partial wave amplitudes, and thus the whole amplitude itself, indeed do not diverge any more for large $\mans$, but rather go to a constant value -- in other words, physical Asymptotic Safety is realised.\footnote{In our setup, we checked that there are admissible gauge choices that implement this, even though they require the fixed point value $g_\ast$ to depend on the gauge parameters in a specific and unexpected way, so we do not report the details.} In such a case, the scalars effectively interact via a $\phi^2\chi^2$-interaction. This is surprising because it looks as if we would break shift symmetry at high energies.

To elaborate, symmetries are conserved along the \ac{RG} flow as long as they are not broken either explicitly (\eg{}~via a symmetry-breaking regulator choice) or spontaneously (\eg{}~via a finite vacuum expectation value). This is because there are modified Ward identities that are preserved by the flow equation. In Asymptotic Safety, the standard conclusion from this is that global symmetries can exist, and all previous computations show this explicitly~\cite{Eichhorn:2011pc, Eichhorn:2012va, Labus:2015ska, Percacci:2015wwa, Eichhorn:2017eht, Eichhorn:2020sbo, Ali:2020znq, deBrito:2021pyi, Eichhorn:2021qet, Laporte:2021kyp, Eichhorn:2025ilu, Assant:2025gto}. By contrast, semi-classical considerations based on black-hole physics and string theory~\cite{Banks:1988yz, Giddings:1987cg, Lee:1988ge, Abbott:1989jw, Coleman:1989zu, Kamionkowski:1992mf, Holman:1992us, Kallosh:1995hi, Banks:2010zn} lead to the no-global-symmetries conjecture, which (if true) constrains the quantum gravity landscape~\cite{Vafa:2005ui, Ooguri:2006in, Brennan:2017rbf, Palti:2019pca, vanBeest:2021lhn, Grana:2021zvf, Agmon:2022thq}. This potential tension between string theory and Asymptotic Safety has been widely discussed, see \eg{}~\cite{Eichhorn:2024rkc, Basile:2025zjc} and references therein.

The above observation however presents a mechanism to bring together no global symmetries at high energies, no symmetry breaking along the flow, and symmetry restoration for low energies -- at least for the case of shift symmetry. The most surprising aspect of this is that it is not even based on black-hole arguments, but simply on the condition of Asymptotic Safety at the level of scattering amplitudes. It is however unclear whether a similar mechanism could be realised for other global symmetries. A different, dynamical mechanism around the no-global-symmetries conjecture has been discussed in~\cite{Borissova:2020knn, Borissova:2024hkc}.

\subsection{Derivative expansion}\label{sec:DE}

We now discuss whether the exact results above can be recovered in a derivative expansion, since resolving the momentum dependence functionally is technically challenging in more realistic scenarios. Concretely, we test whether we can expand the dimensionless form factor in a power series (at least asymptotically), so that
\begin{equation}
    f_t^E(z) \sim \sum_{n\geq0} c_{n,t} z^n \, , \qquad z\to0 \, .
\end{equation}
Due to the simple structure of the flow, we can extract the beta functions for the couplings $c_{n,t}$ and solve them analytically. We have that
\begin{equation}\label{eq:DEflow}
    \dot c_{n,t} - (4+2n) c_{n,t} = 256 \frac{g_t^2}{(n+1)!} \left( -\frac{1}{2} \right)^n \, .
\end{equation}
The solution with the correct boundary conditions (\ie{}~so that couplings reach a fixed point as $k\to\infty$) reads
\begin{equation}
    c_{n,k} = 128 \frac{g_\ast^2}{(n+1)!} \left( -\frac{1}{2} \right)^n \left[ \frac{n+1}{n+2} {}_2F_1(1,n+2,n+3,-\tfrac{g_\ast}{G_N k^2}) - \frac{G_N k^2}{g_\ast + G_N k^2} \right] \, ,
\end{equation}
where we switched back to $k$ for convenience. It is easy to check that in the limit $k\to\infty$, we can sum the series and recover the fixed point solution \eqref{eq:FPLOsol}. From this we learn that to resolve the fixed point, the derivative expansion is reliable in the sense that it converges to the true solution.

Moving on to the \ac{IR}, let us try to extract the physical couplings. For this, we consider the dimensionful quantities and take the limit $k\to0$. For the first three coefficients, we find
\begin{equation}\label{eq:LOtaylor}
\begin{aligned}
    C_0 = \lim_{k\to0} k^{-4} c_{0,k} &= 128 G_N^2 \left( 1 + \ln \frac{G_N k^2}{g_\ast} \right) \, , \\
    C_1 = \lim_{k\to0} k^{-6} c_{1,k} &= \frac{32G_N^2}{k^2} + \frac{32 G_N^3}{g_\ast} \left( 1 + 2 \ln \frac{G_N k^2}{g_\ast} \right) \, , \\
    C_2 = \lim_{k\to0} k^{-8} c_{2,k} &= -\frac{8G_N^2}{3k^4} + \frac{32G_N^3}{3g_\ast k^2} + \frac{16G_N^4}{3g_\ast^2} \left( 1 + 3 \ln \frac{G_N k^2}{g_\ast} \right) \, .
\end{aligned}
\end{equation}
We see that we cannot immediately define the couplings due to \ac{IR} divergences scaling with logarithms or even inverse powers of $k$.\footnote{Note that here, we implicitly think of the $C_i$ as to be measured in Planck units. Even if one would look at dimensionless ratios of only the $C_i$, the limit $k\to0$ would yield nonsensical results. For example, dimensionless ratios of the form $C_n^{1+m/2}/C_m^{1+n/2}$, with $n\neq m$, scale like $k^{2(m-n)}$. For a fixed $m>n$ we would conclude that $C_n=0$ and $C_m\neq0$, but then we can investigate the above ratio with $m\to m'>m$, and $n\to m$, and would now conclude that $C_m=0$ and $C_{m'}\neq0$, ad infinitum, which gives a contradiction.} This would suggest that the effective action would not be well-defined. The phenomenon has been observed recently~\cite{Basile:2021krr, Knorr:2022ilz, Baldazzi:2023pep, Knorr:2024yiu, Eichhorn:2024wba, DelPorro:2025fiu}. Here, we can understand why it happens and whether it poses an issue.

The question of why this happens conceptually is easily answered: in the derivative expansion, we are expanding the action in powers of $P^2/k^2$. This works as long as $P\lesssim k$, and clearly breaks down when we take the limit $k\to0$. From the exact solution \eqref{eq:LO_dimless_solution} we can see explicitly that the limit $k\to0$ and an expansion in powers of $P^2$ do not commute. Likewise, the exact \ac{IR} form factor simply does not admit a Taylor expansion.

We can also pinpoint the source of the issue in the beta functions \eqref{eq:DEflow} directly. If the quantum term were absent, the couplings would scale classically,
\begin{equation}
    c_{n,k} \propto k^{4+2n} \, .
\end{equation}
The problem arises when the quantum term -- in our case $\propto g_k^2$ -- dominates over the classical scaling term at small $k$, and thus
\begin{equation}
    c_{n,k} \propto k^4 \, ,
\end{equation}
independently of $n$. For $n=0$, we find logarithmic running as the quantum term scales like the classical scaling term. For $n>0$ we find the power-law divergences as indicated above for the dimensionful couplings. Clearly, this pattern generalises: whenever the quantum contribution to a beta function competes with or dominates over the classical scaling, cutoff \ac{IR} divergences appear for the corresponding dimensionful coupling. In summary, this general argument suggests that issues must arise for theories with massless fluctuations.\footnote{In theories without massless degrees of freedom, the mass term gaps all logarithms, so that naively a derivative expansion is expected to converge. Indeed, for the three-dimensional Ising model, this is what seems to happen~\cite{Balog:2019rrg}.}

Conceptual issues aside, a relevant question is whether we can still extract the exact solution from the derivative expansion by subtracting the unphysical divergences and identifying the logarithmic $k$-dependence with a logarithmic $P$-dependence. Let us assume this is possible. Starting at $C_0$, we assume that we have to replace
\begin{equation}
    k^2 \to c_{k\to P} P^2 \, ,
\end{equation}
for some constant $c_{k\to P}$ that ideally can be extracted from only the coefficient $C_0$. Equating the exact coefficient \eqref{eq:LOFFexpanded} with $C_0$ in \eqref{eq:LOtaylor}, we find
\begin{equation}
    128 G_N^2 \left( \gamma + \ln \frac{G_N P^2}{2g_\ast} \right) \stackrel{!}{=} 128 G_N^2 \left( 1 + \ln \frac{G_N \, c_{k\to P} P^2}{g_\ast} \right) \, ,
\end{equation}
where $\gamma$ is Euler's gamma constant. While the prefactor of the logarithm comes out right, we need to fix
\begin{equation}\label{eq:scaleID}
    c_{k\to P} = \frac{e^{\gamma-1}}{2}
\end{equation}
for the two couplings to agree. It seems preposterous to us to assume that anyone would identify this as the correct scale identification without the full solution at hand. Nevertheless, let us assume that we would be able to derive this coefficient, and compare $C_1$ to check whether the same scale identification produces the correct coupling. As stated above, the easiest procedure would be to simply subtract any power-law divergent term, and identify logarithms. Doing so yields a difference between the dimensionful couplings of the exact solution and that of the derivative expansion:
\begin{equation}
    C_1^\text{exact} - C_1^\text{DE} = - \frac{32 G_N^3}{g_\ast} \, .
\end{equation}
This is a significant error. This pattern continues for higher-order coefficients: the logarithmic term agrees, but the constant term disagrees. Note also that the disagreement is worse by a factor $\sim2-3$ for the naive scale identification $k^2\to P^2$.

Summarising, our results suggest that \emph{for purely massless theories}, while the derivative expansion seems to be able to reliably identify the logarithmic dependence, it is not suited to compute the constant parts -- \ie{}, the Wilson coefficients -- even under the best possible conditions, for those terms where the quantum fluctuations dominate over the classical scaling term.\footnote{Note that in situations where physical results do not depend strongly on the precise scale identification, scanning a range of different scale identifications~\cite{DelPorro:2025fiu, DelPorro:2025wts} might still allow one to derive robust results in massless theories even within a derivative expansion.} This tends to be the case for classically irrelevant couplings. Notably, we do not expect that including scalar fluctuations changes this picture, since they would not change the dominance of quantum gravity fluctuations over the classical scaling term. We emphasise that such artificial divergences $\sim1/k^2$ must not to be interpreted as physical \ac{IR} non-localities of the form $1/\Box$ or similar. We discuss the practical relevance of these results for theories with massive fluctuations in \Cref{sec:results_massive}.

\subsection{RG improvement}

We now investigate inasmuch standard \ac{RG} improvement techniques can represent the physical momentum dependence qualitatively. For this, we discuss two different ways to perform the \ac{RG} improvement. In the first case, we identify the $k$-dependence of the dimensionful coupling with physical momentum dependence. For the same reasons as discussed in \Cref{sec:DE}, the scale dependence that gets the \ac{IR} value right is \eqref{eq:scaleID}. With this, we would get
\begin{equation}
    \left. F^{E,\text{RGI1}}(P^2) = 128 G_N^2 \left( \frac{1}{1+\frac{G_N k^2}{g_\ast}} - \ln \left( 1 + \frac{g_\ast}{G_N k^2} \right) \right) \right|_{k^2\to c_{k\to P}P^2} \, .
\end{equation}
While this (by construction) gets the small-momentum limit right, the large-momentum limit is qualitatively wrong. Concretely, this form factor falls off too quickly,
\begin{equation}
    F^{E,\text{RGI1}}(P^2) \sim - \frac{256 g_\ast^2 e^{2(1-\gamma)}}{P^4} \, , \qquad P^2 \to \infty \, .
\end{equation}
This is actually an immediate consequence of the mass dimension of the form factor. Consequently, wrong conclusions would be drawn for the high-energy behaviour of scattering amplitudes with this type of \ac{RG} improvement.

\begin{figure}[t]
\centering
\includegraphics[width=.7\linewidth]{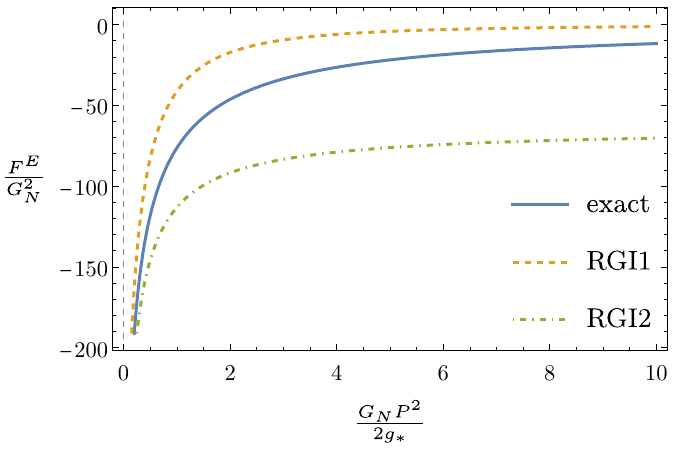} 
\caption{\label{fig:LO_RGI}Comparison of different Euclidean form factors obtained via \ac{RG} improvement to the exact result. We have identified the scale $k$ with the Euclidean momentum $P$ in such a way to exactly match the low-momentum limit, see \eqref{eq:scaleID}. Neither \ac{RG} improvement reproduces the correct large-$P^2$ behaviour.}
\end{figure}

A second possibility is to identify the ratio of dimensionless couplings with the Euclidean form factor at $k=0$ in Planck units,
\begin{equation}
    \left.\frac{c_{0,k}}{g_k^2}\right|_{k^2\to c_{k\to P}P^2} \rightarrow \frac{F^E(P^2)}{G_N^2} \, .
\end{equation}
In this case, we would obtain
\begin{equation}
    \left. F^{E,\text{RGI2}}(P^2) = 128 G_N^2 \left( 1 + \frac{G_N k^2}{g_\ast} \right)^2 \left( \frac{1}{1+\frac{G_N k^2}{g_\ast}} - \ln \left( 1 + \frac{g_\ast}{G_N k^2} \right) \right) \right|_{k^2\to c_{k\to P}P^2} \, .
\end{equation}
While once again the small-momentum regime is recovered correctly by construction, the large-momentum regime is qualitatively incorrect, as this form factor goes to a constant:
\begin{equation}
    F^{E,\text{RGI2}}(P^2) \sim - 64 G_N^2 \, , \qquad P^2 \to \infty \, .
\end{equation}
We show the two results from \ac{RG} improvement together with the exact form factor in \Cref{fig:LO_RGI}. From these examples, we conclude that standard \ac{RG} improvement techniques fail qualitatively in recovering the actual momentum dependence. On top of that, even if they would succeed in recovering the momentum dependence for Euclidean momenta, the extension to Minkowski momenta gives unphysical properties. For example, for the first implementation of \ac{RG} improvement above, we find a \emph{pole} at
\begin{equation}
    P^2 = - \frac{e^{1-\gamma} g_\ast}{G_N} \, ,
\end{equation}
while we find a discontinuity in the imaginary part of the form factor for the second implementation.

\section{Some results for massive theories}\label{sec:results_massive}

The above results are rather concerning, if they were to generalise to all theories with gravitational fluctuations: the general expectation is that quantum gravity effects should be small at observable scales. In particular, they should not completely dominate the physics significantly below the Planck scale, which the logarithms that we found above however do, cf.~\eqref{eq:LOFFexpanded}. In this section, we include some effects of scalar mass terms to understand the situation in massive theories.

\subsection{Setup for massive flows}

Since introducing more parameters complicates the setup significantly, we have to implement additional simplifications. First, we add mass terms for the scalars to the action,
\begin{equation}
    \Gamma_k^{E,M} = \int \text{d}^4x \, \sqrt{g} \left[ \frac{1}{2}M_{\phi,k}^2 \phi^2 + \frac{1}{2}M_{\chi,k}^2 \chi^2 \right] \, .
\end{equation}
For the dimensionless version of the masses, we furthermore approximate the $k$-dependence by a simple scaling,
\begin{equation}
    m_{\phi,k}^2 = M_{\phi,k}^2 k^{-2} \approx m_{\phi,\ast}^2 + M_\phi^2 k^{-2} \, , \qquad m_{\chi,k}^2 = M_{\chi,k}^2 k^{-2} \approx m_{\chi,\ast}^2 + M_\chi^2 k^{-2} \, .
\end{equation}
As before, an asterisk denotes fixed point values, whereas the capital letters without the subscript $k$ indicate the physical masses in the limit $k\to0$.

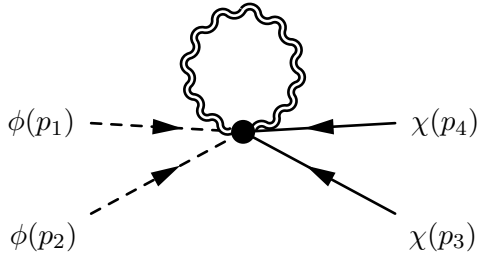
\begin{figure}
\centering
\unitlength = 1mm
\begin{fmffile}{tadpole}
\begin{fmfgraph*}(40,30)
\fmfleft{i1}
\fmfright{o2}
\fmfbottom{i2,o1}
\fmftop{v2}
\fmflabel{$\phi(p_1)$}{i1}
\fmflabel{$\phi(p_2)$}{i2}
\fmflabel{$\chi(p_3)$}{o1}
\fmflabel{$\chi(p_4)$}{o2}
\fmfv{decor.shape=circle,decor.filled=full,decor.size=4thick}{v1}
\fmf{scalar}{i1,v1}
\fmf{scalar}{i2,v1}
\fmf{dbl_wiggly,tension=0.6,left=-0.9}{v1,v2,v1}
\fmf{fermion}{o1,v1}
\fmf{fermion}{o2,v1}
\end{fmfgraph*}
\end{fmffile}

\vspace{0.5cm}
\caption{\label{fig:tadpole}Tadpole diagram that we evaluate additionally for the flow of the form factor $F_{4,k}^E(P^2)$. The wavy double lines represent graviton propagators, whereas the full and dashed lines correspond to the scalar fields. A regulator insertion $k\partial_k \regulatoroperator_k$ is understood on the propagator line.}
\end{figure}

We are now interested in the momentum-dependent generalisation of the $\phi^2\chi^2$ coupling,
\begin{equation}
    \Gamma_k^{E,F_4} = \int \text{d}^4x \, \sqrt{g} \, \frac{1}{4} \phi^2 F_{4,k}^E(\Delta)\chi^2 \, .
\end{equation}
Our model thus probes some structural aspects of, \eg{}, the Higgs quartic coupling. The form factor $F_4$ poses the additional complication that it is dimensionless. This entails that any putative fixed point is necessarily non-local at small momenta, if we only include quantum fluctuations that induce the form factor.\footnote{More explicitly, this can be seen in analogy to \eqref{eq:dimlessflowequationLO} evaluated at the fixed point $\dot f=0$. The difference for $F_4$ is that there is no term proportional to $F_4$ itself, only the $z$-derivative appears. This, together with the observation that the right-hand side is finite at $z=0$, leads to an unphysical logarithmic behaviour, which is lifted as soon as some level of self-feedback is taken into account.} To prevent this truncation artefact, we do not only consider the double seagull diagram shown in \Cref{fig:flow}, but also include an approximation of the tadpole-like diagram, \Cref{fig:tadpole}. Specifically, we extract the four-scalar-two-graviton vertex needed for this diagram only from the volume term $\sqrt{g}$, treating the form factor as metric-independent. In this way, the form factor $F_{4,k}^E$ also appears on the right-hand side of the flow equation, but only as a function of the external momentum, not the loop momentum. Effectively, this term acts as an anomalous dimension. This approximation thus avoids making an integral equation out of the flow equation, which in turn keeps our system solvable analytically. Similar approximations have been successful in computing the momentum dependence of gravitational vertices~\cite{Christiansen:2015rva, Denz:2016qks}.

We emphasise that we still do not take into account scalar loops. In a more complete treatment, we expect that they are crucial for the mechanisms that we discuss below.

\subsection{Momentum-dependent results with masses}\label{sec:massive_mom_dep}

The flow equation for the function $f_{4,t}(z)$ reads, in our approximation,\footnote{Note that since the form factor $F_4$ is dimensionless, the only difference between $F_4$ and $f_4$ is the momentum argument, which is dimensionful in the first case and dimensionless in the second.}
\begin{equation}\label{eq:flow_phi4}
    \dot f_{4,t}^E(z) - 2 z f_{4,t}^{E\prime}(z) = \frac{10 g_t}{\pi} f_{4,t}^E(z) + 320 \frac{1- e^{-\frac{z}{2}}}{z} g_t^2 m_{\phi,t}^2 m_{\chi,t}^2 \, .
\end{equation}
Here, we again used the exponential regulator \eqref{eq:expshapefunction} to evaluate the loop integral. The fixed point solution reads
\begin{equation}
    f_{4,\ast}^E(z) = \frac{80 g_\ast^2 m_{\phi,\ast}^2 m_{\chi,\ast}^2}{1-\frac{5g_\ast}{\pi}} \left[ \frac{2}{z} - \left( \frac{z}{2} \right)^{-\frac{5g_\ast}{\pi}} \left( \Gamma\left(\tfrac{5g_\ast}{\pi}\right) + \left( 1 - \frac{5g_\ast}{\pi} \right) \Gamma\left(\tfrac{5g_\ast}{\pi}-1,\tfrac{z}{2}\right) \right) \right] \, .
\end{equation}
Comparing with \eqref{eq:FPLOsol}, we can see that the inclusion of the tadpole gives rise to an anomalous dimension which dictates part of the large-$z$ behaviour, depending on the size of $g_\ast$,
\begin{equation}
    f_{4,\ast}^E(z) \simeq \frac{80 g_\ast^2 m_{\phi,\ast}^2 m_{\chi,\ast}^2}{1-\frac{5g_\ast}{\pi}} \left[ \left( \frac{z}{2} \right)^{-1} - \Gamma\left(\tfrac{5g_\ast}{\pi}\right) \left( \frac{z}{2} \right)^{-\frac{5g_\ast}{\pi}} \right] \, , \qquad z \to \infty \, .
\end{equation}
For small $z$, we can see the effect of the tadpole diagram even more directly, since
\begin{equation}
    f_{4,\ast}^E(0) = -16\pi g_\ast m_{\phi,\ast}^2 m_{\chi,\ast}^2 \, ,
\end{equation}
that is, the form factor at zero argument is linear in $g_\ast$. This can only come from the tadpole, since the double seagull is quadratic in $g_\ast$.

Moving away from the fixed point, it is actually very difficult to solve \eqref{eq:flow_phi4} exactly for arbitrary values of $g_\ast$. Analytic solutions for all $t$ can be found whenever $g_\ast$ is an integer multiple of $\pi/5$. We can however directly evaluate the limit $k\to0$ for any $g_\ast$ if we treat the limit carefully.\footnote{We verified explicitly for a few examples that the general solution at $k=0$ is consistent with the cases where an explicit solution for all $k$ is available.} The result reads
\begin{equation}\label{eq:f4exact}
\begin{aligned}
    F_4^E(P^2) &= \frac{64 \pi g_\ast^2 m_{\phi,\ast}^2 m_{\chi,\ast}^2}{1 - \left( \frac{5g_\ast}{\pi} \right)^2} \frac{1}{G_N P^2} - \frac{32 g_\ast}{1 + \frac{5g_\ast}{\pi}} \frac{5G_N M_\phi^2 M_\chi^2 + \pi \left( m_{\phi,\ast}^2 M_\chi^2 + m_{\chi,\ast}^2 M_\phi^2 \right)}{P^2} \\
    &\qquad + \frac{160g_\ast^2 \Gamma\left(\tfrac{5g_\ast}{\pi}-1\right)}{G_N P^2} \Bigg[ g_\ast m_{\phi,\ast}^2 m_{\chi,\ast}^2 U\left(\tfrac{5g_\ast}{\pi}-1, -2, \tfrac{G_N P^2}{2g_\ast}\right) \\
    &\hspace{4.5cm} - \frac{1-\frac{5g_\ast}{\pi}}{\pi} \bigg\{ 5 G_N^2 M_\phi^2 M_\chi^2 U\left(1 + \tfrac{5g_\ast}{\pi}, 0, \tfrac{G_N P^2}{2g_\ast}\right) \\
    &\hspace{5cm} + \pi G_N (m_{\phi,\ast}^2 M_\chi^2 + m_{\chi,\ast}^2 M_\phi^2) U\left(\tfrac{5g_\ast}{\pi}, -1, \tfrac{G_N P^2}{2g_\ast}\right) \bigg\} \Bigg] \, .
\end{aligned}
\end{equation}
In this, $U$ is the Tricomi confluent hypergeometric function, which for our case can be defined via
\begin{equation}
    U(a,b,z) = \frac{1}{\Gamma(a)} \int_0^\infty \text{d}t \, e^{-zt}t^{a-1}(1+t)^{b-a-1} \, .
\end{equation}
As in the massless case, \eqref{eq:f4exact} is analytic in the entire complex plane except for the negative real line, where it admits a branch cut. Along this cut,
\begin{equation}
\begin{aligned}
    \text{Im} \, F_4^E(P^2) = \frac{10\pi G_N^2}{3} \Bigg[ &24 M_\phi^2 M_\chi^2 \, {}_1F_1\left( 2 + \frac{5g_\ast}{\pi}, 2, \frac{G_N P^2}{2g_\ast} \right) \\
    &- 6P^2 \left( m_{\phi,\ast}^2 M_\chi^2 + m_{\chi,\ast}^2 M_\phi^2 \right) \, {}_1F_1\left( 2 + \frac{5g_\ast}{\pi}, 3, \frac{G_N P^2}{2g_\ast} \right) \\
    &+ P^4 m_{\phi,\ast}^2 m_{\chi,\ast}^2 \, {}_1F_1\left( 2 + \frac{5g_\ast}{\pi}, 4, \frac{G_N P^2}{2g_\ast} \right) \Bigg] \, , \qquad P^2 \leq 0 \, .
\end{aligned}
\end{equation}

While the exact expression \eqref{eq:f4exact} is not very illuminating, some relevant information lies in its expansion for small $P^2$. We find
\begin{equation}\label{eq:f4exp}
\begin{aligned}
    F_4^E(P^2) \sim &- \frac{16 \pi g_\ast m_{\phi,\ast}^2 m_{\chi,\ast}^2}{1 + \frac{5g_\ast}{\pi}} - \frac{80 g_\ast G_N}{1 + \frac{5g_\ast}{\pi}} (m_{\phi,\ast}^2 M_\chi^2 + m_{\chi,\ast}^2 M_\phi^2) \\
    &\hspace{3cm} + 80 G_N^2 M_\phi^2 M_\chi^2 \left( 2\gamma - 1 + \psi\left(2+\tfrac{5g_\ast}{\pi}\right) + \ln \frac{G_N P^2}{2g_\ast} \right) \\
    & + G_N P^2 \Bigg[ \frac{20 g_\ast m_{\phi,\ast}^2 m_{\chi,\ast}^2}{1 + \frac{5g_\ast}{\pi}} + 10 G_N (m_{\phi,\ast}^2 M_\chi^2 + m_{\chi,\ast}^2 M_\phi^2) \times \\
    &\hspace{5cm} \left( 3 - 4\gamma - 2\psi\left(2+\tfrac{5g_\ast}{\pi}\right) - 2\ln \frac{G_N P^2}{2g_\ast} \right) \\
    &\hspace{1.75cm} + \frac{10 G_N^2 M_\phi^2 M_\chi^2(2 + \frac{5g_\ast}{\pi})}{g_\ast} \left( -5 + 4\gamma + 2\psi\left(3+\tfrac{5g_\ast}{\pi}\right) + 2\ln \frac{G_N P^2}{2g_\ast} \right) \Bigg] \, .
\end{aligned}
\end{equation}
We find logarithmic contributions at each order in the expansion, confirming the pattern observed in the previous section. However, their quantitative relevance has changed. To see this, we inspect the contributions to the ``constant'' part, \ie{}, the first three terms. For finite fixed-point values for the masses, the logarithmic term is exponentially suppressed. To be more precise, we can estimate at which value of $G_N P^2$ the logarithmic term contributes a similar amount as the first term. Up to numerical factors that are unimportant, we find the general order-of-magnitude estimate
\begin{equation}
    G_N P^2 \simeq e^{-1/(G_N^2 M_\phi^2 M_\chi^2)} \, .
\end{equation}
For illustration, if we assume the masses of the scalars to be of the order of the Higgs mass, $M_\phi\simeq M_\chi\simeq 10^2$GeV, we find
\begin{equation}\label{eq:plol}
    P^2 \simeq 10^{-38} e^{-10^{68}} \text{GeV}^2 \, .
\end{equation}
It is clear that this is so absurdly small to be irrelevant in all cases. Consequently, we can safely neglect the logarithm compared to the constant term. The effective momentum dependence for momenta between \eqref{eq:plol} and the scale of the scalar masses is thus constant. Note that the extreme exponential suppression is a consequence of the \emph{prefactor} of the logarithm, $\propto G_N^2 M_\phi^2 M_\chi^2$. This factor is the same suppression expected from \ac{EFT} arguments, which suggest that quantum gravity effects are unimportant for particle physics up to the Planck scale.

The pattern persists for the term linear in $G_N P^2$, with the logarithm being suppressed by another extremely small number. We note however that at this order, also the second term has a logarithm. Starting at order $G_N^2 P^4$, the logarithm appears in all contributions, and thus apparently reintroduces the problem of large gravitational logarithms that are relevant even at the scale of the scalar masses. We believe however that this is an artefact of our truncation, in particular of neglecting scalar loops. Once these are included, following standard \ac{EFT} considerations we expect the leading-order terms not to be suppressed by powers of $(G_N P^2)$, but by powers of $(P^2/M_{\phi/\chi}^2)$, reintroducing the Planck-suppressed prefactor of the logarithms relative to the leading-order term. This argument suggests that in massive theories, the derivative expansion should work extremely well in practice, and that gravitational logarithms are irrelevant. We discuss some potential caveats to this conclusion in the next section.

\section{Consequences: beyond analytically solvable models}\label{sec:consequences}

Let us discuss some implications of our results for general investigations of gravitational \ac{RG} flows. Following the order of the results, we first discuss purely massless theories, followed by theories with massive fields. We then discuss how gravitational couplings are problematic even in massive theories, and then review a practical implementation of resolving functional dependencies.

\textbf{Purely massless theories.}
At the conceptual level, from the arguments discussed in \Cref{sec:DE}, we understand that the minimal requirement to obtain a finite answer for the effective action in massless theories is to have some form of functional dependence on a physical quantity. This can be the momentum dependence of a correlation function, but functions of curvature monomials, field invariants or similar constructions should work as well. An example in pure gravity would be $f(R)$, which has been investigated in great detail in the past, see \eg{}~\cite{Dietz:2012ic, Falls:2013bv, Falls:2014tra, Demmel:2014hla, Ohta:2015efa, Ohta:2015fcu, Demmel:2015oqa, Falls:2016msz, Falls:2017lst, Christiansen:2017bsy, Falls:2018ylp, Alkofer:2018fxj, Alkofer:2018baq, Burger:2019upn, Kluth:2020bdv, Mitchell:2021qjr, Morris:2022btf}. The expectation from our work is that while fixed point studies can be performed successfully in a Taylor expansion, the physical limit $k\to0$ cannot be taken. Rather, the full curvature dependence has to be resolved. Extrapolating our results, our educated guess for the structural form of the function $f(R)$ is
\begin{equation}\label{eq:fR}
    f(R) \sim \frac{\Lambda}{8\pi G_N} - \frac{R}{16\pi G_N} + R^2 \sum_{n\geq 0} \Big[ c_n + \ell_n \ln (G_N R) \Big] \left( G_N R \right)^n \, .
\end{equation}
If qualitatively correct, this might have direct consequences for phenomenological studies in $f(R)$ theories that are motivated by quantum gravity, see \eg{}~\cite{Ben-Dayan:2014isa, Liu:2018hno}. A possible way around the issue is the observation that in many physical scenarios, the curvature is actually finite, and thus the above expansion can be re-expressed as an expansion about a finite value of the curvature. While this saves the analysis on a technical level, the interpretation of the expansion coefficients has to be changed whenever the curvature is not Planckian, that is whenever the logarithms contribute significantly. This reinterpretation might provide an explanation for the very large value of the $R^2$-coupling that is needed in Starobinsky inflation to be compatible with observations~\cite{Planck:2018jri, AtacamaCosmologyTelescope:2025nti}.

\textbf{Massive theories.}
It seems reasonable to assume that as discussed at the end of \Cref{sec:massive_mom_dep}, as soon as massive fluctuations are included, gravitational logarithms are unimportant in practice for all classically irrelevant couplings. Consequently a derivative expansion would work well for all phenomenologically relevant scales. The condition on the classical mass dimension can be motivated through the requirement of having a relative Planck suppression in front of the gravitational logarithm. For relevant couplings (\ie{}~typically masses), we also expect no issues to be present since quantum terms are generically not dominating classical scaling terms.

For marginal couplings, the suppression for the local $\phi^2\chi^2$ interaction found in \Cref{sec:massive_mom_dep} can however be challenged. In Asymptotic Safety, the most prominent fixed point candidate for gravity coupled to the Standard Model has vanishing fixed point masses for all matter fields~\cite{Eichhorn:2012va, Pastor-Gutierrez:2022nki}. While other fixed points might exist, this is the minimal and most investigated scenario. From our expression \eqref{eq:f4exp}, for such a fixed point, the logarithm is actually not suppressed, since the leading-order terms would vanish in this case even if finite masses are generated along the flow. This issue would be relevant in the Standard Model for the Higgs sector. While certainly calling for a more thorough investigation, this issue might put the massless fixed point scenario under some strain. One speculative answer could be the following: as introduced, the scalar masses are related to derivatives of the scalar potential at vanishing field. In the regime of spontaneous symmetry breaking, the effective potential must be convex, and the form factor $F_4$ as well as the derivatives of the potential would have to vanish at zero momentum and field, respectively. This is indeed the case if also the fixed-point values of the masses vanishes, which makes it fit together with the massless fixed point scenario of Asymptotic Safety. The physical masses and couplings would have to be read off from the potential at its finite vacuum expectation value. In brief, the massless fixed point might need spontaneous symmetry breaking to avoid phenomenologically unviable large gravitational logarithms. For a related discussion, see also~\cite{Eichhorn:2017eht}.

The issue of large gravitational logarithms for marginal couplings can likely be avoided for non-minimal fixed points. For example, a fixed point with non-vanishing Yukawa couplings also provides a finite fixed point value for the quartic coupling and the Higgs mass~\cite{Eichhorn:2019dhg, Eichhorn:2021tsx}. For models beyond the Standard Model, a dark sector with a Higgs portal coupling can display a similar mechanism~\cite{Eichhorn:2021tsx}.

\textbf{Gravitational couplings in massive theories.} While we believe that the above discussion applies to matter interactions (and presumably also to non-minimal gravity-matter interactions), for purely gravitational interactions there can still be issues with the derivative expansion even in the presence of mass terms. To illustrate this, let us investigate the one-loop contribution of a free massive scalar field to gravitational operators. Approximating \eqref{eq:Wetterichequation} at one loop order, we have
\begin{equation}
    \Gamma_\Lambda^{1l} - \Gamma_k^{1l} = \frac{1}{2} \text{Tr} \ln \frac{\Delta+M_\phi^2 + R_\Lambda(\Delta)}{\Delta+M_\phi^2 + R_k(\Delta)} \, ,
\end{equation}
where we integrated the flow between an arbitrary \ac{UV} cutoff scale $\Lambda$ and the \ac{IR} scale $k$. The trace can be evaluated with standard heat kernel techniques~\cite{Vassilevich:2003xt, Codello:2012kq}. We are interested in comparing a derivative expansion with the non-local heat kernel (which can capture the full form factors) in the limit $k\to0$. Specifically, we focus on the contributions of the form $R f(\Delta)R$.

The non-local heat kernel gives the expression
\begin{equation}
\begin{aligned}
    \left. \Gamma_\Lambda^{1l} - \Gamma_k^{1l} \right|_{R^2} = \frac{1}{32\pi^2} \, \int \text{d}^4x \sqrt{g} \, R \int_0^{1/4} \text{d}u \, \bigg[ \frac{1}{16} \frac{1}{\sqrt{1-4u}} - \frac{1}{8} &\sqrt{1-4u} - \frac{1}{48} (1-4u)^{3/2} \bigg] \times \\
    &\ln \frac{u\Delta+M_\phi^2 + R_\Lambda(u\Delta)}{u\Delta+M_\phi^2 + R_k(u\Delta)} \, R \, .
\end{aligned}
\end{equation}
Here we used standard formulas~\cite{Knorr:2023usb} and the so-called Ricci-basis of~\cite{Codello:2012kq}. We can now take the limit $k\to0$ and afterwards perform a derivative expansion. For the exponential regulator~\eqref{eq:expshapefunction}, we find in the large $\Lambda$ limit,
\begin{equation}
\begin{aligned}
    \left. \Gamma_\Lambda^{1l} - \Gamma_0^{1l} \right|_{R^2} \sim \frac{1}{32\pi^2} \, \int \text{d}^4x \sqrt{g} \, R \Bigg[ \frac{1}{120} \ln \frac{\Lambda^2}{M_\phi^2} - \frac{1}{336} \frac{\Delta}{M_\phi^2} + \frac{11}{30240} \left( \frac{\Delta}{M_\phi^2} \right)^2 + \dots \Bigg] \, R \\
    + \mathcal O(1/\Lambda^2) \, .
\end{aligned}
\end{equation}
As is expected from \ac{EFT} arguments, every extra power of $\Delta$ is suppressed by the mass of the scalar. By contrast, if we perform a derivative expansion first, we find at finite $k$
\begin{equation}
\begin{aligned}
    \left. \Gamma_\Lambda^{1l,\text{DE}} - \Gamma_k^{1l,\text{DE}} \right|_{R^2} &\sim \frac{1}{32\pi^2} \, \int \text{d}^4x \sqrt{g} \, R \Bigg[ \frac{1}{120} \ln \frac{\Lambda^2}{k^2+M_\phi^2} - \frac{1}{672} \frac{\Delta}{k^2+M_\phi^2} \\
    &\qquad + \frac{11}{362880} \frac{k^2-2 M_\phi^2}{k^2(k^2+M_\phi^2)^2} \Delta^2 + \dots \Bigg] \, R + \mathcal O(1/\Lambda^2) \, .
\end{aligned}
\end{equation}
Here, we can again see issues with the derivative expansion. Not only does the coefficient linear in $\Delta$ come out incorrectly, the higher-order coefficients diverge as $k\to0$. Even if one were to subtract these divergences, the remaining finite parts are incorrect. In brief, we are in a similar situation as in massless theories.

Tracing the origin of this incompatibility reveals that it comes from derivatives of the regulator $R_k$. Clearly, if we first take the limit $k\to0$, the regulator vanishes and cannot induce extra dependencies on $\Delta$. By contrast, if we first expand in powers of $\Delta$, we receive contributions from derivatives of $R_k(\Delta)$ which survive the limit $k\to0$, thereby sourcing the problem. In particular, all terms except the logarithm are regulator-dependent (in contradistinction to first taking the limit $k\to0$, which is regulator-independent), so that arbitrary results can be generated.\footnote{We find it plausible that this issue is (at least in part) caused by the background field formalism that implicitly underlies the construction, see the discussion in~\cite{Pawlowski:2020qer, Pawlowski:2023gym}. Specifically, the kinetic term of the scalar depends on the full metric, whereas the regulator can only involve the background metric, not the fluctuating graviton. The unphysical background field dependence in the regulator can even affect the existence of fixed points~\cite{Bridle:2013sra}.} From this example and our results in \Cref{sec:results}, we can only call for caution when using the derivative expansion in combination with a background field approximation to investigate the limit $k\to0$ in quantum gravity computations, even when massive fields are part of the theory. We emphasise that these technical issues come on top of the conceptual issue of potentially large gravitational logarithms discussed above. Specifically, we expect that the $R^2$-term shares the fate of the quartic scalar coupling. As we speculated above, the solution might be similar in spirit as well: a finite vacuum expectation value for the curvature.

\textbf{Numerics.} Last but not least, we discuss how a practical implementation is able to converge to the correct result in situations where analytical control cannot be achieved. For this, we suggest a two-step strategy. In the first step, for large $k$ the \emph{dimensionless} flow equation is solved on a grid, \ie{}~with fixed grid points measured in units of $k$, \eg{}~$z_i=P_i^2/k^2$. At some finite transition scale $k=k_0$, we switch to dimensionful quantities, \ie{}, we rescale everything by the appropriate power of $k_0$. Step two consists in solving the \emph{dimensionful} flow for $k\in[0,k_0]$ on a grid, \ie{}~with fixed grid points in units of $k_0$, \eg{}~$Z_i=P_i^2/k_0^2$. This ensures that we probe finite physical momenta even in the limit $k\to0$.\footnote{An alternative was used in \cite{Silva:2024wit}, where the variable was made dimensionless with the help of the running gravitational constant. While this would also work in our specific case, more generally such a procedure fails whenever $g_k$ is not a monotonic function of $k$. The latter is the case in many computations, and in general it is not known a priori whether $g_k$ is indeed monotonic.} It is also clear that in the deep \ac{IR}, when $k$ is close to zero, we probe arbitrarily high ratios $P^2/k^2$, so that a high resolution is needed. Different high-accuracy methods have been implemented within the context of \ac{FRG} computations, see \eg{}~\cite{Borchardt:2015rxa, Borchardt:2016pif, Grossi:2019urj, Sattler:2024ozv}.

\section{Conclusion and outlook}\label{sec:conclusion}

In this paper, we have computed a two-to-two scalar scattering amplitude in Asymptotic Safety from first principles. For this, we partially resolved the momentum dependence of a scalar four-point function induced by quantum gravity fluctuations. We first focussed on a shift-symmetric theory. The whole computation could be performed analytically, which for the massless theory led us to the following conclusions:
\begin{itemize}
    \item Asymptotic Safety provides unambiguous and finite results for the effective action, parameterised in terms of its relevant couplings,
    \item quantum-gravitational fluctuations induce large gravitational logarithms that dominate the \ac{IR} of massless theories,
    \item global shift symmetry is broken effectively at high energies at the level of scattering amplitudes in Asymptotic Safety, realising a version of the no-global-symmetries conjecture,
    \item an \ac{RG} fixed point found with respect to a fiducial scale does not necessarily imply physical Asymptotic Safety at the level of amplitudes,
    \item the derivative expansion fails quantitatively to represent physical momentum dependence, but succeeds in finding \ac{RG} fixed points.
\end{itemize}
We then extended our analysis to also include some effects of mass terms for the scalar fields, and found that most, but crucially not all, issues of massless theories are parametrically avoided as long as there is a scale separation between the Planck scale and the mass scale of the scalar. In particular, the derivative expansion works at an effective level in most cases. One exception that we identified are classically marginal couplings, which can still show large gravitational logarithms that dominate the \ac{IR} physics.

We think that many of these observations generalise beyond our simple system and approximation. Most importantly, we insist that Asymptotic Safety should predict a finite effective action devoid of divergences in the fiducial scale $k$, and any such divergence must be an indication of an insufficient approximation. We also believe that in realistic, gapped theories that have a scale separation from the Planck scale (like the Standard Model), large gravitational logarithms should not dominate physics at phenomenologically relevant scales. For classically marginal couplings, we hypothesized that this issue could be solved by a finite vacuum expectation value for the curvature in the gravitational case, and enforced symmetry breaking for scalar fields. For purely gravitational couplings, we however found indications that even in gapped theories, the derivative expansion can fail to yield finite predictions for the effective action.

If these lessons indeed hold true generally, future investigations that aim to investigate the physical limit $k\to0$ in theories that contain gravity need to maintain a functional dependence on either momentum, curvature, or fields (or even combinations thereof), otherwise they risk making qualitatively or quantitatively wrong predictions. \ac{RG} improvement is not a valid simplification technique, since at least in our case it failed qualitatively to represent the momentum dependence.

To gain a more reliable understanding of scattering processes in Asymptotic Safety, it is clearly mandatory to improve the approximation and to discuss other matter content that is more realistic. While the momentum dependence of propagators is by now under control~\cite{Christiansen:2012rx, Codello:2013fpa, Christiansen:2014raa, Meibohm:2015twa, Knorr:2019atm, Bosma:2019aiu, Bonanno:2021squ, Knorr:2021niv, Fehre:2021eob, Knorr:2023usb, Pastor-Gutierrez:2024sbt, Kher:2025rve, Pawlowski:2025etp}, less is known about interaction vertices~\cite{Christiansen:2015rva, Denz:2016qks, Christiansen:2017cxa, Knorr:2017fus, Eichhorn:2018akn, Eichhorn:2018ydy, Eichhorn:2018nda, deBrito:2025nog}, for recent progress on the scalar-scalar-graviton vertex see~\cite{Chiesa:2026tlz}. Resolving their momentum dependence is crucial not only to establish physical Asymptotic Safety, but also to understand whether the theory is unitary and causal. Moreover, matter fluctuations have to be taken into account. Another restriction that needs to be lifted is that of using a flat background. It seems reasonable to expect that some of the problems of massless theories disappear when curvature dependence is included, whenever the curvature acts as an effective mass for the massless fluctuations.

A different handle on unitarity and causality could come from positivity and causality bounds derived within \ac{EFT}~\cite{deRham:2022hpx}. Standard implementations however rely on the physics that has been integrated out to have a mass gap, so they are strictly speaking inapplicable in their original form, as has been noted several times, see \eg{}~\cite{Alberte:2020jsk, Herrero-Valea:2020wxz, Alberte:2021dnj, Knorr:2024yiu, Eichhorn:2024wba}. Our results re-emphasise the need to extend these bounds to properly include massless quantum fluctuations.

\paragraph{Acknowledgements}

I would like to thank Astrid Eichhorn, Zois Gyftopoulos, Marc Schiffer, and Fabian Wagner for interesting discussions, as well as Astrid Eichhorn and Jan Pawlowski for constructive feedback on an earlier draft of this work.

\appendix

\section{NLO}

\begin{figure}[t]
\centering
\includegraphics[width=.7\linewidth]{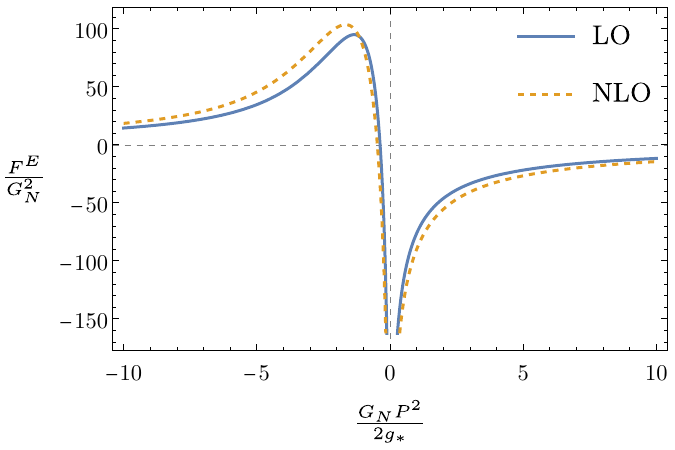} 
\caption{\label{fig:LO_vs_NLO_IR}Comparison of the real parts of the Euclidean form factor at leading and next-to-leading order. Since we have a completely analytical expression, we can evaluate the form factor also for negative arguments. The qualitative momentum dependence is the same, and only small quantitative differences appear.}
\end{figure}

In this appendix we briefly present the results on the form factor in the massless theory when we include the anomalous dimension of the Newton's constant within the beta function of the form factor. In our setup, we have
\begin{equation}
    \eta_{N,k} = -\frac{2g_k}{g_\ast} \, .
\end{equation}
Compared to the leading-order result in the main text, we thus get an additional $g_k^3$-term on the right-hand side. The complete solution for the dimensionless form factor can be expressed as
\begin{equation}
\begin{aligned}
    f_k^{E,\text{NLO}}(z) &= \frac{64 g_\ast (G_N k^2)^2 e^{-\frac{2z}{3}}}{(g_\ast+ G_N k^2)^2 z} \left( 3G_N k^2 - 4 e^{\frac{z}{6}} (g_\ast + 2 G_N k^2) + e^{\frac{2z}{3}} (4g_\ast + 5 G_N k^2) \right) \\
    &\qquad + \frac{128 g_\ast^2 (G_N k^2)^2}{(g_\ast + G_N k^2)^2} \left( \Gamma(0,\tfrac{2z}{3}) - \Gamma(0,\tfrac{z}{2}) \right) \\
    &\qquad - 256 e^{\frac{G_N k^2 z}{2g_\ast}} (G_N k^2)^2 \left( \Gamma(0,\tfrac{G_N k^2 z}{2g_\ast}) - \Gamma(0,\tfrac{(g_\ast + G_N k^2)z}{2g_\ast}) \right) \\
    &\qquad + 128 e^{\frac{2G_N k^2 z}{3g_\ast}} (G_N k^2)^2 \left( \Gamma(0,\tfrac{2G_N k^2 z}{3g_\ast}) - \Gamma(0,\tfrac{2(g_\ast + G_N k^2)z}{3g_\ast}) \right) \, .
\end{aligned}
\end{equation}
This gives rise to the fixed point solution
\begin{equation}
\begin{aligned}
    f_\ast^{E,\text{NLO}}(z) = &\frac{32 g_\ast^2}{z^2} e^{-\frac{2z}{3}} \left( 9 - 6z + e^{\frac{2z}{3}}(23-12z) + 8 e^{\frac{z}{6}} (z-4) \right) \\
    &- 128 g_\ast^2 \left( \Gamma(0,\tfrac{z}{2}) - \Gamma(0,\tfrac{2z}{3}) \right) \, .
\end{aligned}
\end{equation}
Once again, the \ac{IR} quantity $G_N$ drops out as needed. Lastly, the physical dimensionful form factor reads
\begin{equation}
    F^{E,\text{NLO}}(P^2) = -128 G_N^2 e^{\frac{G_N P^2}{2g_\ast}} \left( 2 \Gamma(0,\tfrac{G_N P^2}{2g_\ast}) - e^{\frac{G_N P^2}{6g_\ast}} \Gamma(0,\tfrac{2G_N P^2}{3g_\ast}) \right) \, .
\end{equation}
We compare the real parts of the leading-order with the next-to-leading order result in \Cref{fig:LO_vs_NLO_IR}. We find that qualitatively, both form factors agree. Both have the same scaling for small and large momenta, and only differ in the numerical prefactors. Interestingly, we can write
\begin{equation}
    F^{E,\text{NLO}}(P^2) = 2F^{E,\text{LO}}(P^2) - F^{E,\text{LO}}(\tfrac{4P^2}{3}) \, ,
\end{equation}
where $F^{E,\text{LO}}$ is the leading-order form factor as given in the main text, \eqref{eq:FsolLOIR}. This relation makes it straightforward to perform the same analysis about small and large momentum limits as well as the branch cut, so we will not spell out the details. When once again comparing to the \ac{EFT}, we can extract the refined cutoff scale
\begin{equation}
    \Lambda_\text{EFT}^2 = \frac{8g_\ast}{3G_N} e^{-(1+\gamma)} \, .
\end{equation}
This is larger than the leading-order result \eqref{eq:EFTcutoffLO} by a factor of $4/3$.

Finally, the attentive reader might wonder in which sense the solution above is next-to-leading order -- after all, this is not obvious from the expression in the limit $k\to0$. There are two indications for that. First, in a regime where $g_k$ is small, the $\eta_{N,k}$-term adds a $g_k^3$-contribution to the leading-order $g_k^2$-contribution. Second and somewhat contrary, taking the limit of large fixed point values $g_\ast\to\infty$ has both the leading-order and the next-to-leading order solution diverge logarithmically with $g_\ast$. However, there difference is finite,
\begin{equation}
    \lim_{g_\ast\to\infty} \left[ F^{E,\text{NLO}}(P^2) - F^{E,\text{LO}}(P^2) \right] = -128 G_N^2 \ln \frac{4}{3} \, .
\end{equation}
These two observations lead to the suggestion that the $\eta_{N,k}$-term is sub-leading in a large range of theory space.

\bibliographystyle{JHEP}
\bibliography{bibliography.bib}

\end{document}